\documentclass[twocolumn]{aastex61}
\usepackage{natbib, color,amsmath}
\usepackage[normalem]{ulem}
\usepackage{epstopdf}
\usepackage{graphicx}
\definecolor{red}{rgb}{1.0,0.0,0.0}
\definecolor{white}{rgb}{1.0,1.0,1.0}

\newcommand{\Mj}[1]{$M_\mathrm{Jup}$}

\usepackage[utf8]{inputenc}

\begin{document}
\title{BAFFLES: Bayesian Ages for Field Lower-Mass Stars}

\correspondingauthor{S Adam Stanford-Moore}
\email{astanfordmoore@gmail.com}

\author[0000-0002-0689-2607]{S Adam Stanford-Moore}
\affiliation{Kavli Institute for Particle Astrophysics and Cosmology, Stanford University, Stanford, CA 94305, USA}

\author[0000-0001-6975-9056]{Eric L. Nielsen}
\affiliation{Kavli Institute for Particle Astrophysics and Cosmology, Stanford University, Stanford, CA 94305, USA}
\affiliation{Department of Astronomy, New Mexico State University, P.O. Box 30001, MSC 4500, Las Cruces, NM 88003, USA}

\author[0000-0002-4918-0247]{Robert J. De Rosa}
\affiliation{Kavli Institute for Particle Astrophysics and Cosmology, Stanford University, Stanford, CA 94305, USA}
\affiliation{European Southern Observatory, Alonso de C\'{o}rdova 3107, Vitacura, Santiago, Chile}

\author[0000-0003-1212-7538]{Bruce Macintosh}
\affiliation{Kavli Institute for Particle Astrophysics and Cosmology, Stanford University, Stanford, CA 94305, USA}

\author[0000-0002-1483-8811]{Ian Czekala}
\altaffiliation{NASA Hubble Fellowship Program Sagan Fellow}
\affiliation{Department of Astronomy, 501 Campbell Hall, University of California, Berkeley, CA 94720-3411, USA}


\begin{abstract}

Age is a fundamental parameter of stars, yet in many cases ages of individual stars are presented without robust estimates of the uncertainty. We have developed a Bayesian framework, {\tt BAFFLES}, to produce the age posterior for a star from its calcium emission strength (log($R'_{HK}$)) or lithium abundance (Li EW) and $B-V$ color.  We empirically determine the likelihood functions for calcium and lithium as functions of age from literature measurements of stars in benchmark clusters with well-determined ages. We use a uniform prior on age which reflects a uniform star formation rate. The age posteriors we derive for several test cases are consistent with literature ages found from other methods. {\tt BAFFLES} represents a robust method to determine the age posterior probability distribution for any field star with $0.45 \leq B-V \leq 0.9$ and a measurement of $R'_{HK}$ and/or $0.35 \leq B-V \leq 1.9$ and measured Li EW.  We compile colors, $R'_{HK}$, and Li EW from over 2630 nearby field stars from the literature and present the derived {\tt BAFFLES} age posterior for each star.

\end{abstract}

\section{Introduction}

Age, along with mass and metallicity, is a fundamental parameter of stars. Accurate stellar ages are needed in a wide variety of astronomical studies, including galactic evolution, globular clusters, open clusters, star forming regions, stellar multiples, brown dwarf companions, and planetary systems. For direct imaging exoplanet surveys, such as the Gemini Planet Imager Exoplanet Survey \citep{macintosh:2018,nielsen:2019}, stellar age is important at all stages of the survey.  First, while selecting target stars, younger stars are preferred, since their planets will be inherently brighter and easier to detect.  Second, the mass for an imaged planet is derived from the age of the host star using evolutionary models that link mass, age, and luminosity (e.g. \citealt{allard:2014, baraffe:2015}), and the dominant measurement uncertainty in deriving mass is from age \citep{bowler:2016}.  Third, age is a requirement for measuring the occurrence rates of planets.  Translating sensitivity in apparent brightness to mass sensitivity requires the age of each observed star. Thus completeness to planets as a function of mass, a key ingredient for occurrence rate, relies heavily on precise ages for the entire sample (e.g. \citealt{bowler:2016,nielsen:2013,nielsen:2019}). 

 Yet unlike the ages of stars in co-eval groups, like open clusters or moving groups, the ages of field stars are difficult to determine robustly. For stellar clusters with well-determined membership lists, the main sequence turnoff is used to robustly determine the age (e.g. \citealt{Goudfrooij:2014msto,Cummings:2018msto}). The lithium depletion boundary is applicable to both clusters and more sparse moving groups, with the reddest objects in an association with detectable lithium absorption setting the overall age (e.g. \citealt{burke:2004LDB,soderblom:2010}). 
 For isolated field stars, however, a less robust set of observables that track age are available, including spectroscopic indicators (e.g. \citealt{skumanich:1971,wright:2004}), gyrochronology (e.g. \citealt{kraft:1967,Barnes:2009gyro}), and asteroseismology (e.g. \citealt{Cunha:2007astero}). Here we present a Bayesian method to determine age through two spectral indicators: calcium emission strength and the depth of the lithium absorption line.

\subsection{Empirical Age Indicators}
\subsubsection{Calcium Emission Strength}
Calcium emission strength, as given by the index $R'_{HK}$, is connected to the strength of a star’s magnetic field through the stellar dynamo \citep{noyes:1984}.  The rotation of the star and convection within induces a magnetic field whose strength is proportional to the rate of rotation \citep{noyes:1984,skumanich:1971}. Over time the star's rotation inevitably slows as it ejects ionized particles in its stellar wind, which carry away angular momentum \citep{kraft:1967,weber:1967}, and as a result the magnetic field strength and thus calcium emission strength generally decrease with age.

The index $R'_{HK}$ is a measure of the flux in the narrow emission line in the core of the Calcium II H and K absorption lines at $\sim$3968 \AA\ and $\sim$3934 \AA\ respectively \citep{noyes:1984,wright:2004}. $R'_{HK}$ is derived from an intermediate index, the S index, which represents the ratio of the narrow emission flux to the background continuum flux.  S provides a relative comparison of emission strength, yet includes both chromospheric and photospheric contributions and is dependent on $B-V$ (as well as age).  Therefore, to remove the dependencies on $B-V$ the S index is transformed by two empirically determined polynomials in $B-V$, resulting in $R'_{HK}$ \citep{noyes:1984,wright:2004}, where the polynomials have been calibrated over a $B-V$ range of 0.45 to 0.90, corresponding to an approximate spectral type range of F6 to K2. In addition to the long-term decline in activity over time, the S value for a single star also varies by $\sim$10\% over that star's activity cycle \citep{wright:2004}.

\subsubsection{Lithium Equivalent Width}

The strength of the lithium absorption line traces the amount of lithium present in the photosphere of a star.  When stars initially form, their primordial lithium abundances are similar to the abundance from Big Bang nucleosynthesis, with number densities of $\sim$10$^{-9}$ that of hydrogen \citep{sestito:2005}. Over time stars deplete their primordial lithium by nuclear burning in the core and convective mixing, so that measurements of remaining surface lithium correlate with stellar age \citep{soderblom:2010,skumanich:1971}. For stars cooler than $\sim$7000 K, lithium abundance can be measured based on the equivalent width (EW) of the absorption of the lithium doublet at 6708 \AA\  \citep{soderblom:2010}; hotter stars (OBA spectral types) have ionized their lithium, and have negligible 6708 \AA\ absorption even with no lithium burning. 

Lithium's two isotopes, $^6$Li and the more abundant $^7$Li, burn at temperatures of 2.2 million K and 2.6 million K, respectively.  Since stellar surface temperatures are much lower ($\sim$2500 K for low-mass M stars and $\sim$46000 K for high-mass O-stars), in order to burn, lithium must be brought into hotter layers via convection \citep{soderblom:1990}.  As a result, the rate of lithium depletion largely depends on the depth of the convection zone, allowing lower-mass stars -- which while having lower surface temperatures have much deeper convective layers -- to deplete lithium faster than higher-mass stars \citep{soderblom:1990}.  In addition to convection it is thought that slow mixing induced by rotation and angular momentum loss may affect lithium depletion \citep{sestito:2005}, so that lithium abundance is a function of age, spectral type, and the initial rotation rate and rotational evolution of the star.

\subsection{Functional fits to $R'_{HK}$ and Li EW evolution}\label{sec:previous_fits}

Previous studies have taken advantage of the correlation between $R'_{HK}$ and age to create empirical fits of mean cluster $\log(R'_{HK})$ vs. log cluster age \citep{soderblom:1991,Donahue:1993,Lachaume:1999, mamajek:2008}.  However while these polynomial fits allow one to find an expression for age as a function of $R'_{HK}$, the polynomial makes no direct prediction of uncertainty in the age derivation.  \citet{soderblom:1991} found the standard deviation of their stellar data around their power-law fit to be $\sim$0.2 dex and concluded that ages predicted from their fit would be accurate to $\sim$50\%.  A similar approach has been done for lithium as well (finding an average fit to clusters, and assigning a single age to a star based on its location relative to the cluster fits), e.g. \citealt{mamajek:2002,nielsen:2010}.  However this method fails to capture the full astrophysical scatter.   

In addition, many polynomial fits fail to account for a uniform star formation rate in the Milky Way (with exceptions such as the second polynomial fit developed by \citealt{soderblom:1991}).    Both \citet{soderblom:1991} and \citet{mamajek:2008} note that using a 1-to-1 polynomial conversion between $R'_{HK}$ and age on a volume-limited sample of solar-type stars results in an unphysically large number of stars with ages $<$1 Gyr, compared to older stars, inconsistent with the expected local star formation history (their Figures 9 and 14, respectively).  The polynomial fits (e.g. Figure~\ref{fig:ca_vs_age}) tend to have slopes that become more negative when going to increasing age, so that the curve is flatter at small ages and steeper at larger ages.  If scatter in $R'_{HK}$ is symmetric (which we present evidence for in Section~\ref{sec:ca_scatter}), this leads to a bias where systematically younger ages are predicted, since a 0.1 dex displacement toward more positive values of $R'_{HK}$ moves along the flatter part of the curve to much younger ages, compared to an equal 0.1 dex displacement toward more negative values of $R'_{HK}$, which moves along the steeper part of the curve, and does not move toward older ages as quickly.  \replaced{We simulate this effect by using}{To illustrate this effect we use} the \citet{mamajek:2008} polynomial giving log(age) as a function of $R'_{HK}$ (their Equation 3).  We generate $10^6$ stars uniformly distributed in age between 1 Myr and 10 Gyr, then numerically invert the polynomial to assign a value of $R'_{HK}$ to each, add 0.1 dex of Gaussian noise to each value, and use the polynomial to convert back to age.  The returned age distribution has a significant spike at $\lesssim$1 Gyr, which becomes more prominent as the amplitude of the Gaussian noise is increased.  \citet{soderblom:1991} attempted to correct for this effect by adjusting the polynomial fit at large ages, by constraining it with the nearby star sample and assuming that sample had a uniform star formation rate.  \citet{mamajek:2008} advocates for an activity/rotation/age relation instead, which flattens out the age distribution of the volume-limited sample out to 6 Gyr.  Here, we present an explicit prior uniform in age when creating age posterior probability density functions to address this issue.

Furthermore, the median age estimates for separate methods (e.g. $R'_{HK}$ and Li EW) are difficult to rigorously combine without precise uncertainty estimates.  Previous works have, for example, simply averaged the ages obtained from $R'_{HK}$ and lithium e.g. \citealt{nielsen:2010}.

\citet{brandt:2014_method} develops a Bayesian method to combine the age PDF of a star’s likely moving group with its posterior PDF from indicators of chromospheric and X-ray activity and stellar rotation. The two age distributions are then averaged, weighted by the probability of membership to the moving group. Other works (e.g. \citealt{Casagrande:2011}, \citealt{nielsen:2013}) have developed Bayesian methods for deriving age posteriors from isochrones that also utilize a uniform star formation rate prior.     

We describe here a method to derive Bayesian ages for field stars from lithium or calcium measurements, Bayesian Ages For Field LowEr-mass Stars ({\tt BAFFLES}).\footnote{Our {\tt BAFFLES} package is available at \url{https://doi.org/10.5281/zenodo.3840244} and can be used from the command line with \textit{python baffles.py -bmv [B-V] -rhk [Log($R'_{HK}$)] -li [Li EW]} (with other options available).} For calcium emission our method is calibrated to stars with $B-V$ between 0.45 and 0.9 ($\sim$F6--K2) and log($R'_{HK}$) between $-3.7$ and $-5$.  For lithium we have calibrated {\tt BAFFLES} to stars with $B-V$ between 0.35 and 1.9 ($\sim$F2-M5) and Li EW between 3.2 and 1500 m\AA.

\section{Data}

\begin{table*}[h!]
\centering
\begin{tabular}{c c c c c c c} 
\hline\hline
Group Name & Age (Myr) & Age Ref. & $N_{\rm Ca}$ & Ca Ref. & $N_{\rm Li}$ & Li Ref. \\
\hline
NGC2264	& 5.5 & 3 &	& & 123 & 9,10 \\
Upper Scorpius & 10 & 20 & 8 & 1 & &  \\
UCL+LCC	& 16 & 21,22 & 8 &	1 & & \\
$\beta$ Pic & 24 & 2 & 6 & 1,30,31 & 37 & 14,19 \\
IC2602 & 43.7 &	4 & & & 27 & 11 \\
Tuc/Hor & 45 & 2 & 6 & 1,32,33 & & \\
$\alpha$ Per &	85 & 1,23,24 & 12 & 1 & 60 & 15	\\
Pleiades & 130 & 1,23,25 & 42 & 1 & 128 & 6 \\
M35 & 200 & 26 & & & 82 & 16 \\
M34 & 240 & 5 &	& & 49 & 12 \\
UMa	& 500 &	27 & 10 & 1 & & \\
Coma Ber & 600 & 13 & & & 13 & 17 \\
Hyades	& 700 & 28,34 & 41 & 1 & 50	& 7 \\
M67	& 4000 & 29,18 & 70	& 1 & 40 & 8 \\ 
\hline
\end{tabular}
\caption{{\tt BAFFLES} benchmark clusters for both calcium emission and lithium abundance.  $N_{\rm Ca}$ and $N_{\rm Li}$ refer to the number of stars from each cluster with literature calcium/lithium measurements. References: (1) \citet{mamajek:2008}, (2) \citet{bell:2015},  (3) \citet{turner:2012}, (4) \citet{randich:2018}, (5) \citet{Meibom:2011}, (6) \citet{soderblom:1993pleiades}, (7) \citet{pace:2012}, (8) \citet{jones:1999}, (9) \citet{tobin:2015}, (10) \citet{king:1998}, (11) \citet{randich:2001}, (12) \citet{jones:1997}, (13) \citet{king:2005}, (14) \citet{mentuch:2008}, (15) \citet{Balachandran:2011}, (16) \citet{anthonyTwarog:2018}, (17) \citet{ford:2001},(18) \citet{VandenBerg:2004}, (19) \citet{shkolnik:2017}, (20) \citet{Pecaut:2016_usage}, 
(21) \citet{mamajek:2002}, (22) \citet{deZeeuw:1999}, (23) \citet{barrado:2004}, (24) \citet{Makarov:2006}, (25) \citet{duncan:1991}, (26) \citet{sung:1999}, (27) \citet{king:2003}, (28) \citet{brandt:2015hyades-age}, (29) \citet{Giampapa:2006}, (30) \citet{wright:2004}, (31) \citet{gray:2006}, (32) \citet{jenkins:2006}, (33) \citet{henry:1996}, (34) \citet{gossage:2018hyades-age}.  }
\label{table:clusters}
\end{table*}

We calibrate {\tt BAFFLES} using benchmark moving groups and open clusters with well-determined ages. While calcium emission strength and lithium abundance serve as indicators of relative age, we use these clusters to calibrate the relationships that give age as a function of indicator.  Table \ref{table:clusters} gives basic properties on each benchmark cluster as well as our assumed age for each.

Since the ages of the benchmark clusters anchor the calcium and lithium age relations, accurate ages are important to the accuracy of {\tt BAFFLES}; the offset in the posteriors scales with the factor by which the ages are modified. For both lithium and calcium, modifying the age of a single cluster by $\pm 1\sigma$, tends to change the median age derived by {\tt BAFFLES} by $\lesssim 3\%$.  Systematically shifting all the cluster ages in the same direction by $\pm 1\sigma$ shifts the derived median ages of posteriors by a comparable amount, $\lesssim 20\%$.  

\subsection{Calcium benchmark clusters}

Ages, stellar $R'_{HK}$ values, and stellar $B-V$ values used in this work for calcium were compiled by \citet{mamajek:2008}, though here separately reported measurements for the same star are averaged together (though this had little effect on our fits). Adopted ages for the benchmark clusters were mostly identical to those adopted by \citet{mamajek:2008}, except we used more recent age estimates of 24 Myr for $\beta$ Pic and 45 Myr for Tuc/Hor from \citet{bell:2015}, 10 Myr for Upper Scorpius from \citet{Pecaut:2016_usage}, and $\sim$700 Myr for Hyades from \citet{brandt:2015hyades-age,gossage:2018hyades-age}.   

\subsection{Lithium benchmark clusters}
We compiled $B-V$ and Li EW measurements from the multiple sources listed in Table \ref{table:clusters}. For duplicate stars we averaged measurements, and used the measurement if there was one measurement and one upper limit.

We used the stellar $B-V$ values if they were provided for individual stars; otherwise we used $B-V$ magnitudes \replaced{from Simbad (for nearby associations with negligible reddening) or }{compiled from the literature in Table \ref{table:B_V_ref}, which are all in nearby moving groups with negligible reddening. For stars in clusters with significant reddening, we }converted spectral type or $T_{eff}$ to $B-V$.  \citet{soderblom:1993pleiades} (Pleiades), \citet{jones:1997} (M34), and \citet{jones:1999} (M67) reported dereddened $(B-V)_0$, while \citet{randich:2001} (IC2602), \citet{anthonyTwarog:2018} (M35), \citet{ford:2001} (Coma Ber), and \citet{pace:2012} (Hyades) gave uncorrected $B-V$. NGC2264 lithium equivalent widths from \citet{tobin:2015} were not accompanied by $B-V$ values, so we converted spectral type to $B-V$ using the conversion in \citet{pecaut:2013}. For $\alpha$ Per \citep{Balachandran:2011}, we converted $T_{eff}$ (which had been inferred from $V-K$ color) to $B-V$ also using the conversion in \citet{pecaut:2013}.  For $\beta$ Pic \citep{mentuch:2008,shkolnik:2017}, a moving group $\lesssim$100 pc, we expect negligible reddening, and we took the observed $B-V$ colors \replaced{(from Simbad)}{(given in Table \ref{table:B_V_ref})} to be the intrinsic colors.

Close binaries present an issue since it is not always clear whether the $B$ magnitude, $V$ magnitude, or lithium absorption are resolved or from the combined systems. To avoid this issue, for $\beta$ Pic moving group members we removed binaries from \citet{mentuch:2008}: AZ Cap, CD-64 1208, GJ 3305, AT Mic A, AT Mic B, HIP 23418, LP 476-207 and binaries from \citet{shkolnik:2017}: PM J01071-1935, LP 467-16, Barta 161 12, BD+17 232, CD-44 753, PM J05243-1601, GSC 06513-00291, MCC 124, TWA 22, CD-64 1208, AT Mic, GR* 9. We also removed stars with poorly measured values of $B$ or $V$ magnitudes (uncertainty $\gtrsim$0.15 mags) from \citet{mentuch:2008}: HD 164249B and from \citet{shkolnik:2017}: FK Psc, BD+30 397, EXO 0235.2-5216, 2MASS J05200029+0613036, RX J0520.5+0616, Smethells 20, CD-31 16041, TYC 6872-1011-1, TYC 7443-1102-1, BD-13 6424, UCAC4 396-055485.          

For M67, we removed stars identified by \citet{jones:1999} as being less secure members, as well as potentially unresolved binaries. In many cases Li EW is given without measurement error, with the exception of \citet{mentuch:2008} and \citet{randich:2001}, which did provide individual errors. For Coma Ber, we also omitted stars \citet{ford:2001} identified as non-members or spectroscopic binaries.

\section{Methods}

{\tt BAFFLES} is a Bayesian framework which finds a star's posterior age probability density function (PDF) from input of $R'_{HK}$, or $B-V$ combined with Li equivalent width, or all three.  We calibrate the method using datasets of the benchmark clusters discussed above.

\subsection{Calcium}

Using the cluster data, we first present an age posterior from an $R'_{HK}$ measurement of calcium emission.

\subsubsection{Framework}

We seek an expression that returns an age PDF for a single star given an $R'_{HK}$ measurement, which is the posterior

\begin{equation}
    p(t|\hat{r})
\end{equation}

\noindent where $t$ is the age and $\hat{r}$ is the measured value of $R'_{HK}$ for a single star, with measurement uncertainty of $\sigma_{\hat{r}}$.  We evaluate this posterior using Bayes' rule

\begin{equation}
    p(\theta | D) = \frac{1}{Z} p(D | \theta) p(\theta)
\end{equation}

\noindent where the four terms are posterior ($p(\theta | D)$), evidence ($Z$), likelihood ($p(D | \theta)$), and prior ($p(\theta)$), functions of the data ($D$) and parameters of the model ($\theta$).

\added{For calcium the parameters of our model, $\theta$, are the age $t$ and the true value of $R'_{HK}$ for the star, $r$, while the data, $D$, are our measured value of $R'_{HK}$, $\hat{r}$. We also assume the evidence, $Z$, is a constant.  With these terms Bayes' rule becomes  

\begin{equation}
    p(r,t | \hat{r}) \propto p(\hat{r} | r,t) p(r,t).
\label{eq:calcium_bayes}
\end{equation}
}

\replaced{For Calcium, our}{Our} knowledge of the true value of $R'_{HK}$ for the star, $r$, comes from a measurement with an associated measurement error: $\hat{r}$ and $\sigma_{\hat{r}}$.  In the case of $R'_{HK}$ the astrophysical scatter among stars in a single cluster is generally much larger than the measurement \added{uncertainty} for any one star.  Thus, our model should incorporate both the overall trend that for clusters of different ages, \deleted{that}average $r$ ($\mu_r$) decreases with increasing age, and that there is a scatter about this mean at a single age ($\sigma_r$).  We expect both these terms to evolve with time, and express them as functions $\mu_r = f(t)$ and $\sigma_r = g(t)$.  If the scatter is fit by a Gaussian, our prior on $r$ then becomes

\begin{equation}
\label{eq:calcium_gaussian}
    p(r|t) = \mathcal{N}(r | f(t), g(t))
\end{equation}

\noindent while the prior on t, $p(t)$, is flat for a uniform star formation rate, uniform in linear age between 1 Myr and 13 Gyr.  Although the star formation rate increases at ages older than $\sim$8 Gyr, this prior is a reasonable approximation for ages $<$ 5 Gyr \citep{snaith:2015}, which also corresponds to the oldest benchmark clusters we utilize. Higher-mass stars have main sequence lifetimes shorter than the full range of our prior. A stellar lifetime prior is a complicated function of $B-V$, especially since stars of a given mass evolve in color over time. Rather than commit to a particular set of isochrones, we choose to keep {\tt BAFFLES} as empirically-driven as possible. An isochrone-based age prior can be applied to a {\tt BAFFLES} posterior once generated, and we advise caution when considering an age posterior with significant probability at very large ages for higher-mass stars.  Together, these define a joint prior for our problem

\begin{equation}
    p(r,t) = p(r|t) p(t) \added{= \mathcal{N}(r | f(t), g(t))p(t)}.
\end{equation}

\replaced{
The likelihood of our data is assumed to be a Gaussian,

\begin{equation}
    \mathcal{L}(\hat{r} | r) = \mathcal{N}(\hat{r} | r, \sigma_{\hat{r}})
\end{equation}

\noindent if the measurement error, $\sigma_{\hat{r}}$, is significant.  

We have no direct data on the age, $t$, but it is a parameter of our model, so we rewrite the likelihood as 

\begin{equation}
    \mathcal{L}(\hat{r} | r) = \mathcal{L}(\hat{r} | r, t) = p(\hat{r} | r,t).
\end{equation}

\noindent Then, from Bayes' rule, we can write the posterior as

\begin{equation}
    p(r,t | \hat{r}) = \frac{p(\hat{r} | r,t) p(r,t)}{Z}
\label{eq:calcium_bayes}
\end{equation}

\noindent where $Z$ is the evidence, which we assume to be a constant.

We simplify this expression by assuming that $\sigma_{\hat{r}}$ is negligible, especially given the larger astrophysical scatter, $\sigma_r$, and instead take the likelihood to be a delta function, 

\begin{equation}
   p(\hat{r} | r, t) = \mathcal{L}(\hat{r} | r) = \delta(r - \hat{r}).
\end{equation}
}{

In the general case of measurements with Gaussian error bars, likelihood would be given by a normal distribution,

\begin{equation}
    \mathcal{L}(\hat{r} | r) = \mathcal{N}(\hat{r} | r, \sigma_{\hat{r}}).
\end{equation}

\noindent However, for $R'_{HK}$ we assume that the uncertainty is negligible, especially given the larger astrophysical scatter, $\sigma_r$. Therefore we instead take the likelihood to be a delta function,

\begin{equation}
   p(\hat{r} | r) = \mathcal{L}(\hat{r} | r) = \delta(r - \hat{r}).
\end{equation}

We have no direct data on the age, $t$, but it is a parameter of our model, so we rewrite the likelihood as 

\begin{equation}
    p(\hat{r} | r) = p(\hat{r} | r,t) = \delta(r - \hat{r}).
\end{equation}

\noindent We can now rewrite Equation \ref{eq:calcium_bayes}, the joint posterior over $r$ and $t$, as

\begin{equation}
    p(r,t | \hat{r}) \propto \delta(r - \hat{r})\mathcal{N}(r | f(t), g(t))p(t)
\end{equation}

and after marginalizing over $r$ and taking $p(t)$ to be a constant, we solve for $p(t|\hat{r})$,

\begin{equation}
    p(t | \hat{r}) \propto \int p(r,t | \hat{r}) dr = \mathcal{N}(r | f(t), g(t)).
\label{eq:ca_posterior_end_of_framework}
\end{equation}

}

\noindent \replaced{Thus, once we determine functional forms for $f(t)$ and $g(t)$ from our cluster data, we can evaluate the likelihood and produce a posterior for any star given a measurement $\hat{r}$.}{If the astrophysical scatter is Gaussian, then by determining functional forms for $f(t)$ and $g(t)$ from our cluster data we can evaluate the likelihood and produce a posterior for any star given a measurement $\hat{r}$. In Section \ref{sec:ca_scatter}, however, we present evidence that the scatter is not well-modeled by a Gaussian, and introduce a new numerical function to describe the prior on r.}

\subsubsection{The Color Dependence of $R'_{HK}$}

\begin{figure*}
  \centering
  \begin{tabular}{c @{\qquad} c }
    \includegraphics[page={1},width=.48\linewidth]{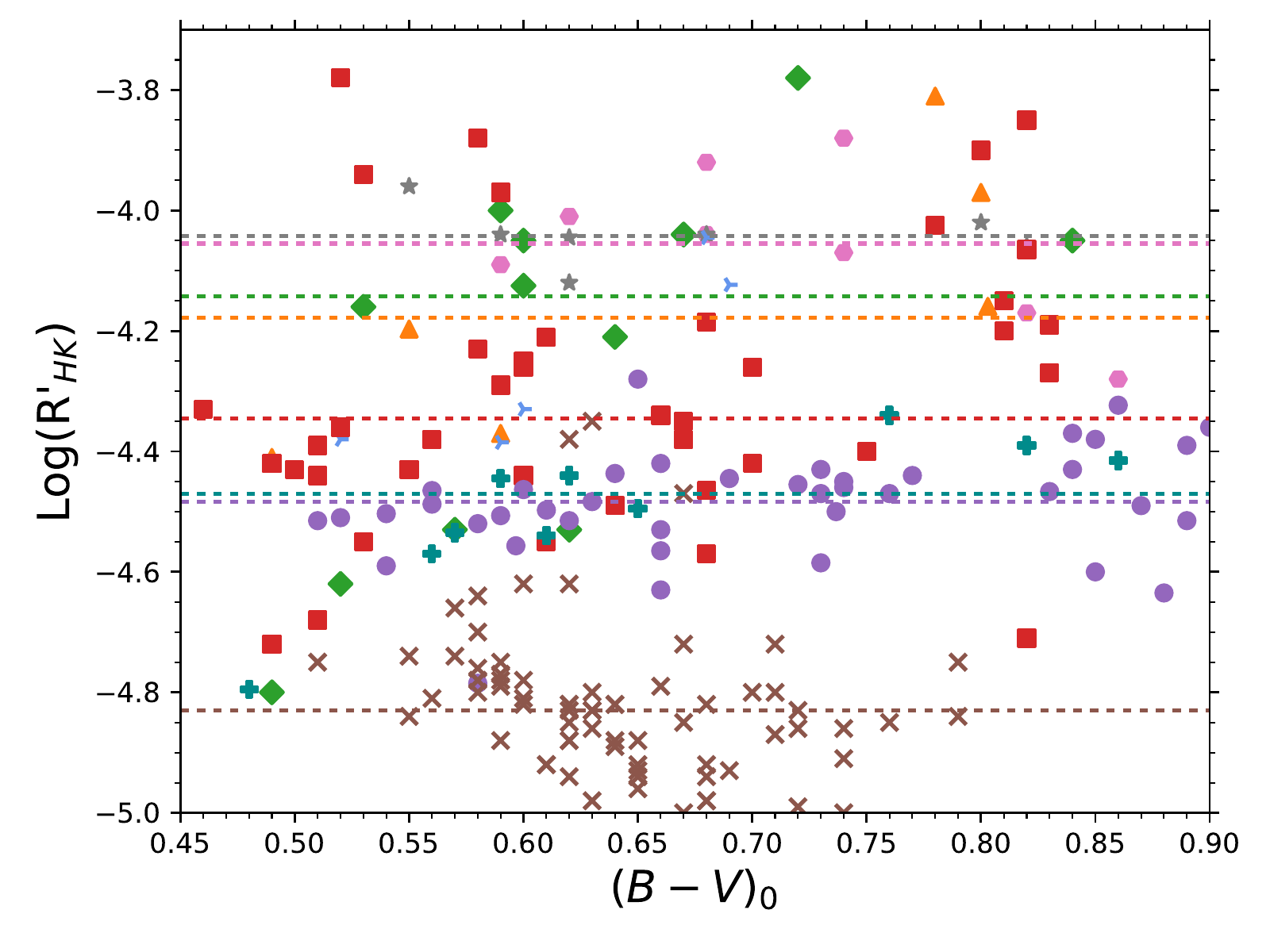} &
    \includegraphics[page={2},width=.48\linewidth]{calcium_metal_vs_bv.pdf} \\
    \small (a) One-parameter fits & \small (b) Two-parameter fits
  \end{tabular}
  \caption{(left) One-parameter fits with no color dependence avoid overfitting. (right) linear fits to the $B-V$ dependence of $R'_{HK}$ for each cluster are dominated by outliers for sparse datasets, and become non-monotonic at the blue and red ends.  We adopt the one-parameter fit in the final prior function $f(t)$. }
  \label{fig:linear-constant-fits}
\end{figure*}

The derived quantity $R'_{HK}$ is formulated to be independent of $B-V$ color, which is accomplished by using two polynomials in $B-V$ to convert the raw $S_{HK}$ value into $R'_{HK}$.  To determine the extent to which $R'_{HK}$ is in fact independent of color we initially considered using a two-parameter linear fit to the cluster \added{$R'_{HK}$ as a function of $B-V$} (similar to \citet{mamajek:2008}), since the slopes seemed non-negligible.  However, since our dataset included many clusters with only a handful of calcium measurements, the fit slopes were poorly determined, and the fits crossed frequently, a non-physical outcome.   As in the right panel of Figure \ref{fig:linear-constant-fits}, linear fits to the clusters resulted in non-monotonic changes in $R'_{HK}$ over time, especially in the reddest and bluest regions of our $B-V$ range.  Although \citet{mamajek:2008} used linear fits for each cluster, they interpolated cluster means for solar $B-V$ ($\sim$0.65) only.  For solar $B-V$, the cluster means are still monotonic, something not true for other $B-V$ values that were included in our study.

There is a significant improvement in $\chi^2$ from the linear fit to the constant fit, dropping from 554 (constant) to 423 (linear), assuming a constant measurement error for each star of 0.1 dex, as estimated by \citealt{mamajek:2008}, which from the Bayesian information criterion presents very strong evidence in favor of the linear model ($\Delta$BIC = 83.3).  Nevertheless, we find the behavior of the linear fits in the right panel of Figure~\ref{fig:linear-constant-fits} to be unphysical, where at the reddest and bluest ends the evolution in $R'_{HK}$ is non-monotonic, and implies wild swings in calcium activity as a function of age, based on a handful of datapoints in each cluster, and poor sampling across the entire $B-V$ range.  As a result, to avoid over-fitting sparse data, we adopted a constant fit for $R'_{HK}$, where each cluster is represented by the median value of $R'_{HK}$, with no $B-V$ dependence. A constant fit has the advantage of capturing the monotonic decrease in $R'_{HK}$ while remaining the simplest fit. \citet{mamajek:2008} advocate determining age from $R'_{HK}$ through an age-activity-rotation relation, the effect of which is a significant $B-V$ dependence on $R'_{HK}$ for objects of similar ages (see their Figure 11), which varies by $\sim$0.15 dex across $B-V$. As there are limited $R'_{HK}$ measurements in benchmark clusters it is currently difficult to confirm this behavior of $R'_{HK}$ as a function of color.  In fact, more direct solutions to a $B-V$ dependence of $R'_{HK}$ would be to either redetermine the polynomial parameters or to fit directly in $S_{HK}$, and either would likely require a larger dataset than that presented here.

\subsubsection{$R'_{HK}$ as a function of age}

\begin{figure}[htp]
\centering
\includegraphics[width=\columnwidth]{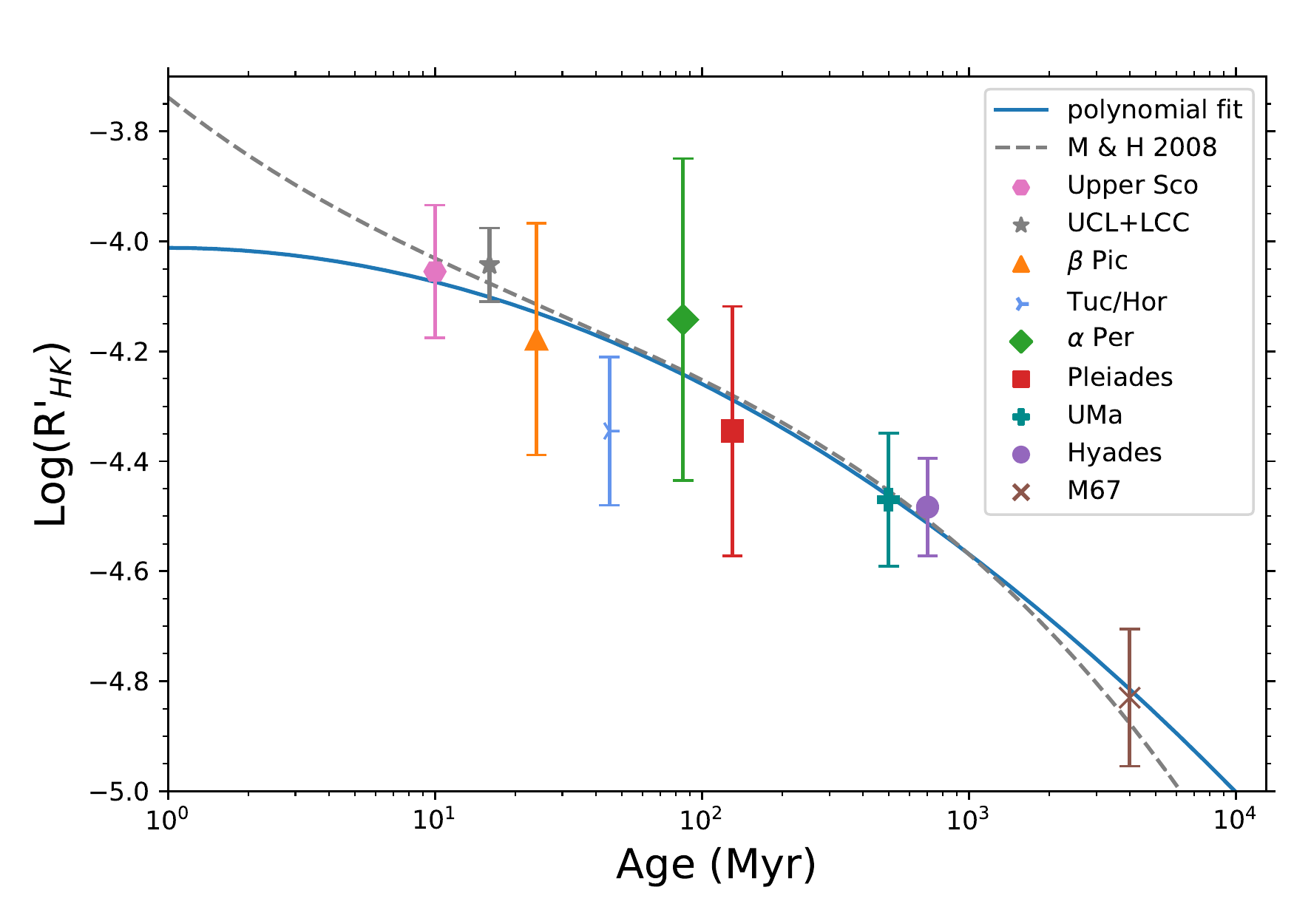}
\caption{Fit for the mean $R'_{HK}$ as a function of age using cluster median values. Error bars on each point represent the standard deviation of $\log(R'_{HK})$ in each cluster, while the fit is weighted only by the number of points in each cluster.  The observed variation is consistent with $R'_{HK}$ decreasing monotonically with time. The solid blue line is our second-order polynomial fit to cluster median activity, which is very similar to the overplotted third-order polynomial (gray dashed line) from \citet{mamajek:2008}, based on a nearly identical dataset.}
\label{fig:ca_vs_age}
\end{figure}

From the fits above we have nine cluster ages and their respective mean log($R'_{HK}$) values, which we use to find the mean log($R'_{HK}$) at all ages covered by our prior, $\mu_r = f(t)$. We fit log($R'_{HK}$) as a function of age with a second-order polynomial, constrained to be monotonically decreasing, and where each cluster in the fit is weighted by the number of stars it contains.  Figure \ref{fig:ca_vs_age} shows this fit against the median value of each cluster, with plotted error bars indicating the standard deviation in each cluster.  Our fit is consistent with polynomial fits from previous authors.  \citet{mamajek:2008} use linear fits for finding each cluster’s mean $R'_{HK}$ as a function of $B-V$, and then fit a third-order polynomial to age, based on the value of each cluster's linear fit evaluated at solar $B-V$ of 0.65. The largest discrepancies between the two fits are, unsurprisingly, at ages lower than that of the youngest benchmark cluster (Upper Sco) and larger than that of the oldest (M67). \citet{soderblom:1991} experimented with several different second-order polynomials, correcting for disk heating and a uniform star formation rate.

\subsubsection{Astrophysical Scatter}\label{sec:ca_scatter}

\begin{figure}[htp]
\centering
\includegraphics[width=\columnwidth]{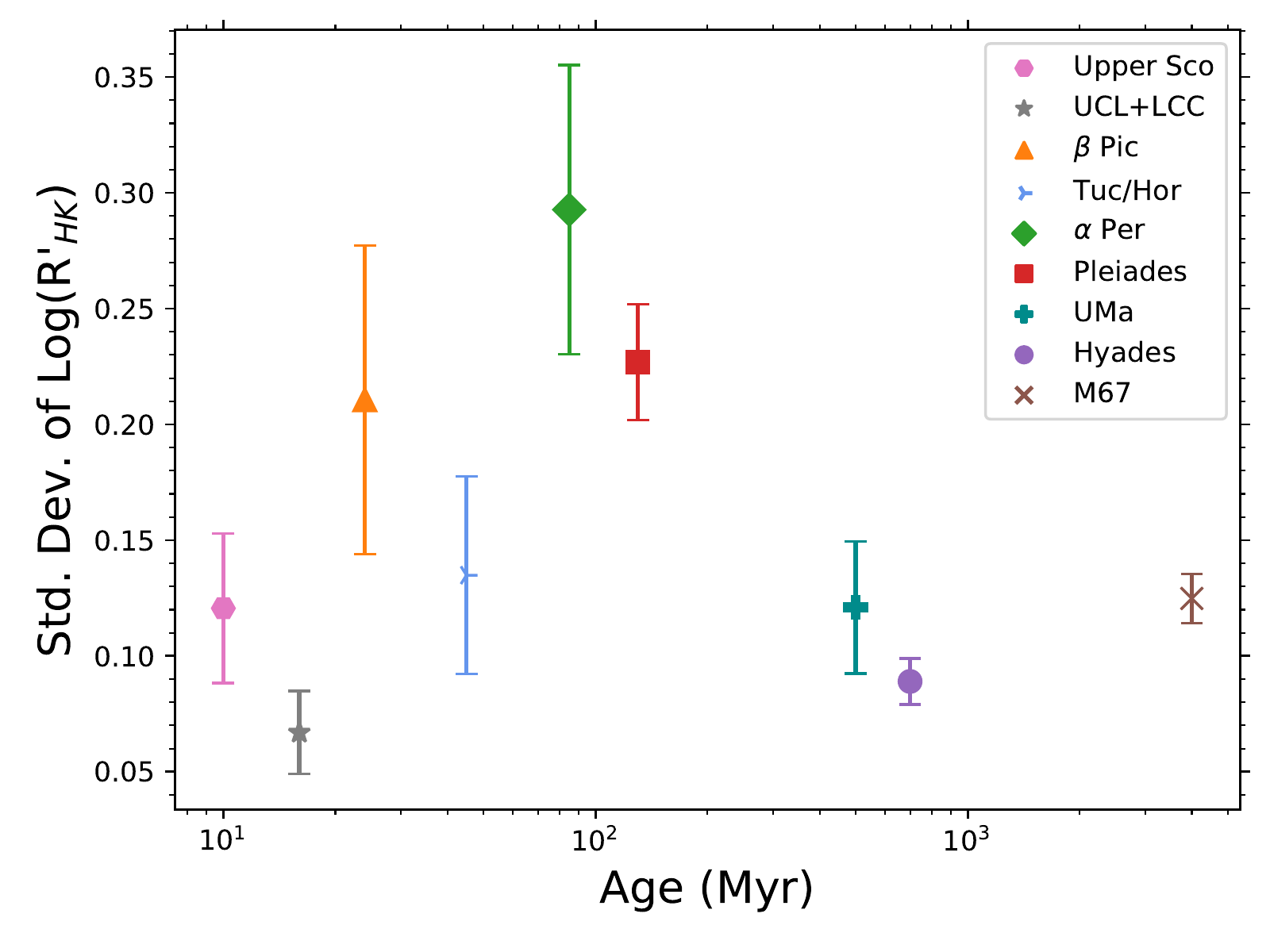}
\caption{Standard deviation\added{s} of the residuals to the median of each calcium cluster\replaced{, computed as the standard deviation of the residuals}{, and computed uncertainties}.  Scatter in $R'_{HK}$ over time appears to increase between 20 and 200 Myr, suggestive of a similar effect in the scatter of rotation rate as a function of age \citep{gallet:2013}. \replaced{As a result we model the scatter as a function of age as a Gaussian function in log(Age) with parameters $\mu_S$=1.96, $\sigma_S$=0.16, amplitude $A_S$=0.077, and constant offset $C_S$=0.11.  There are limited numbers of young stars with $R'_{HK}$ measurements, (8 stars each in Upper Sco and UCL+LCC, and 6 stars each in $\beta$ Pic and Tuc/Hor), making our measurements of standard deviation somewhat uncertain at these ages, compared to clusters with more measurements.}{While the three clusters with the most measurements, Pleiades, Hyades, and M67 show a significant offset at $\sim$100 Myr compared to $>$700 Myr, many of the remaining clusters have fewer than 10 stars with calcium measurements, and as such poorly determine the standard deviation. More measurements of stars in these young clusters are needed to robustly map out any age dependence.}}
\label{fig:calcium_scatter_vs_age}
\end{figure}

We next examine the astrophysical scatter of $R'_{HK}$ about the mean, $\sigma_r = g(t)$.  We begin by computing the residuals of $R'_{HK}$ for \replaced{each}{every} star in a cluster to the median value for all stars in the cluster.  The standard deviations of these residuals \replaced{is}{are} plotted in Figure~\ref{fig:calcium_scatter_vs_age}\replaced{. }{, where uncertainty in the standard deviation ($\sigma_m$) of the m'th cluster with $N_m$ stars is given by the equation appropriate for gaussian scatter, $\pm \frac{\sigma_m}{\sqrt{2N_m - 2}}$.}  \replaced{We observe}{There is some evidence} that the scatter between 20-200 Myr is larger than the scatter for younger or older stars. This is reminiscent of Figure 1 of \citet{gallet:2013}, where solar-type stars spin up between $\sim$10-50 Myr as they contract when approaching the main sequence, and the dispersion in rotation rate between the fast-rotators and slow-rotators in a single cluster increases, compared to stars younger than 20 Myr or older than 200 Myr.  \replaced{As a result we fit a four-parameter Gaussian distribution to the standard deviation of the residuals as a function of age, shown as the dashed orange line in  \ref{fig:calcium_scatter_vs_age}.  This fit represents standard deviation as a function of time, 

\begin{equation}
    \sigma_r = g(t) = \frac{A_S}{\sigma_S \sqrt{2 \pi}} e^{-\frac{1}{2}\left(\frac{log(t) - \mu_S}{\sigma_S} \right)^2} + C_S
\label{eq:sigma_r_gaussian}
\end{equation}

\noindent with fit parameters of amplitude ($A_S$), constant ($C_S$), mean ($\mu_S$), and standard deviation ($\sigma_S$).  While this Gaussian does appear to be a reasonable fit to the data, we don't have strong evidence that this is the correct functional form to describe the evolution of calcium emission over time.  Looking ahead, larger datasets of more stars in these young clusters will be needed to better define this effect.}{We investigated using a Gaussian or inverted parabola to fit the data, but we don't have strong evidence that such a fit is justified.  The clusters with the most measurements, Pleiades, Hyades, and M67, show the strongest evidence for a change in the standard deviation with time. However, the clusters Upper Sco, UCL+LCC, $\beta$ Pic, Tuc/Hor, $\alpha$ Per, all have few stars (8, 8, 6, 6, and 12 respectively), making the standard deviations not well-determined.  As a result we treat g(t) as a constant, and note that larger datasets with more stars in these young clusters would be needed to precisely measure time-dependent behavior. } 

\begin{figure*}
\centering
  \begin{tabular}{c @{\qquad} c }
    \includegraphics[page={1},width=.48\textwidth]{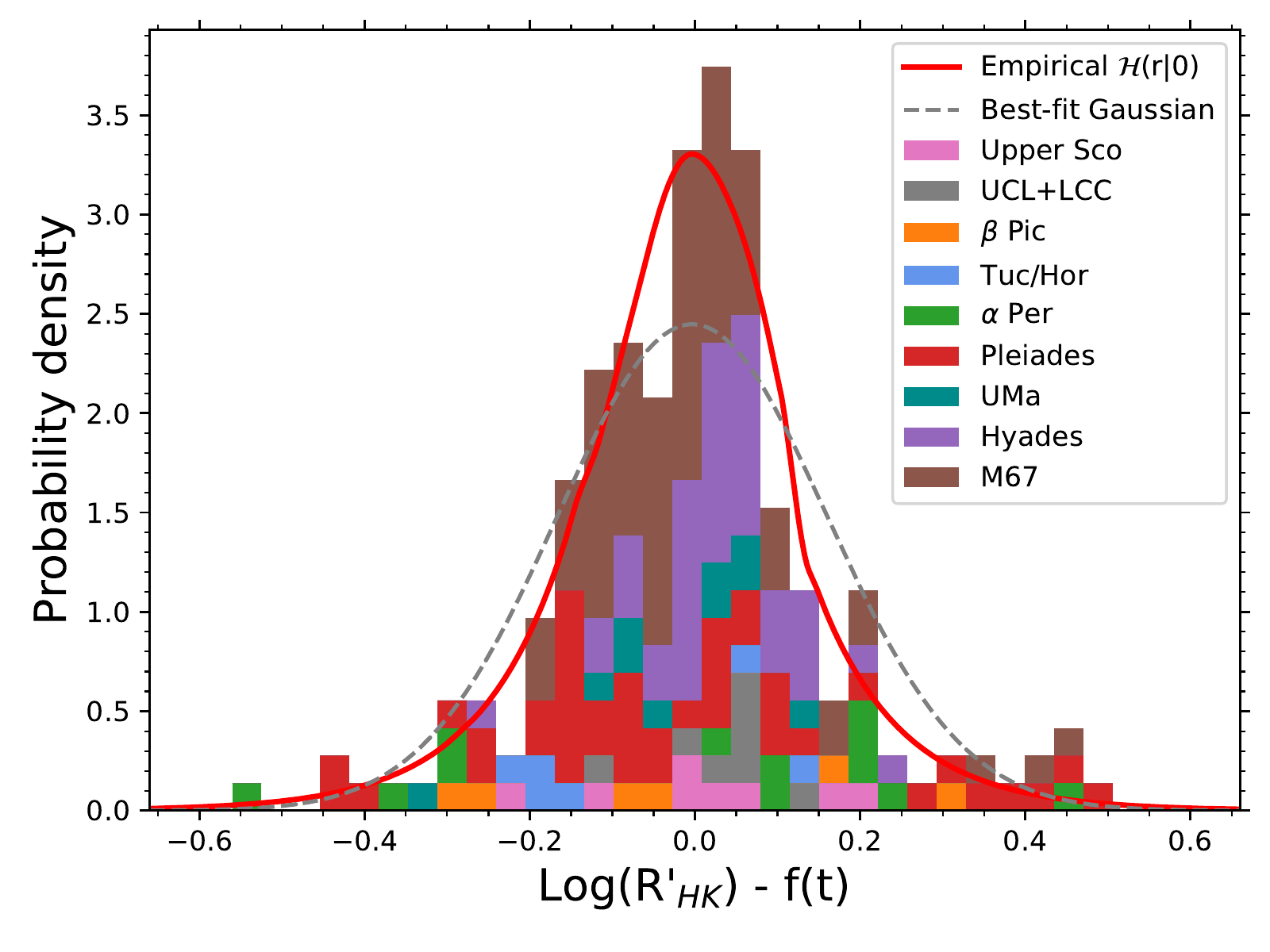} &
    \includegraphics[page={2},width=.48\textwidth]{calcium_hist_fit.pdf} \\
  \end{tabular}
\caption{(left) A histogram of residuals of all cluster stars to \replaced{the median $R'_{HK}$ value of each cluster ($\mu_m$), divided by the fit to the standard deviation ($g(t)$), $a=\frac{\hat{r} - \mu_m}{g(t)}$.}{the fit $f(t)$.}  The distribution appears significantly non-Gaussian, as the best-fit Gaussian (dashed gray line) has a lower peak and overpredicts the number of stars at $\sim$1$\sigma$.  Instead, we construct a numerically determined function from our data (red line, \replaced{$h(r|f(t),g(t))$}{${\cal H}(r|f(t))$}).  (right) CDF of the residuals, from which we construct the empirical function \replaced{$h$}{${\cal H}$}. }
\label{fig:calcium_pdf}
\end{figure*}

\replaced{Plotting the residuals to the cluster median ($\mu_m$) in Figure~\ref{fig:calcium_pdf} for every star in the cluster, and dividing by the fit standard deviation from Equation~\ref{eq:sigma_r_gaussian}, $a = \frac{\hat{r} - \mu_m}{g(t)}$, we}{We instead compute the residuals for each star with respect to the fit $f(t)$, and} observe that these residuals, while somewhat symmetric, are not well-fit by a Gaussian (the dashed gray curve \added{in Figure \ref{fig:calcium_pdf}}).  In particular, the best-fit Gaussian underestimates the peak and is significantly wider at $\sim$1$\sigma$ compared to the data.  We thus return to Equation~\ref{eq:calcium_gaussian}, and replace the normal distribution in our prior with a new \replaced{empirical}{numerical} function, 

\replaced{
\begin{equation}
    p(r|t) = \replaced{h}{{\cal H}}(r | f(t),g(t)).
\end{equation}
}{
\begin{equation}
    p(r|t) = \replaced{h}{{\cal H}}(r | f(t))
\end{equation}
}

\noindent which has a mean $f(t)$\replaced{, and width set by the standard deviation, $g(t)$.}{. The amplitude of the scatter is encoded by ${\cal H}$ itself, and since we take $g(t)$ to be a constant, the shape of ${\cal H}$ does not change over time, only its mean.}

To evaluate \replaced{$h$}{${\cal H}$}, we fit a function to the smoothed CDF of the \deleted{scaled}residuals, as in the right panel of Figure \ref{fig:calcium_pdf}, capturing the non-Gaussian shape of the astrophysical scatter. The tails of the PDF are constrained to decrease exponentially out to \replaced{5}{$\sim$4} standard deviations and then are fixed at zero.  We perform the smoothing with a Savitzky-Golay filter, which fits successive windows with a 3rd-order polynomial, so as to remove jumps in the function from star to star, but without significantly increasing the width of the distribution, and the final function is a good fit to the data (left panel of Figure~\ref{fig:calcium_pdf}).  We normalize the final \replaced{$h$}{${\cal H}$} distribution so it has an integral of unity. \added{Then, since we find ${\cal H}$ to be slightly asymmetric with a median of 0.0026 dex, we shift the distribution so that its median has a value of zero}.  \deleted{As \replaced{$h$}{${\cal H}$} is derived from the residuals divided by $g(t)$, the scatter in $r$ at a single age $t$ increases from $\sim$20-200 Myr compared to younger and older ages, as we observe in our cluster dataset.}

\added{The numerical fit ${\cal H}$ is one of many possible implementations of the \replaced{likelihood}{prior} function. Other choices with wider tails (such as the Student's-t distribution or the Lorentzian distribution) also partially capture the non-Gaussian behavior. We found the best fit with the Student's-t distribution, which came closest to matching the residual distribution, and found no significant difference between it and the empirical function ${\cal H}$ on our final age posteriors.}

\subsubsection{Calcium posterior}

Rewriting Equation~\ref{eq:calcium_bayes}, we then have an expression for our posterior given by 

\replaced{
\begin{equation}
    p(r,t|\hat{r}) \propto \delta(r - \hat{r}) h(r | f(t), g(t)) p(t)
\end{equation}
}{
\begin{equation}
    p(r,t|\hat{r}) \propto \delta(r - \hat{r}) {\cal H}(r | f(t)) p(t)
\end{equation}
}

\noindent where $f$\deleted{, $g$,} and \replaced{$h$}{${\cal H}$} are determined from our cluster datasets above, and p(t) is constant for a uniform star formation rate.  \replaced{We find a posterior only in age by marginalizing over $r$, which is greatly simplified given our delta function likelihood,}{We can then rewrite our calcium age posterior in equation \ref{eq:ca_posterior_end_of_framework} using our function ${\cal H}$ as} 

\replaced{
\begin{equation}
  p(t | \hat{r}) = \int p(r,t | \hat{r}) dr \propto h(\hat{r} | f(t), g(t)).
\label{eq:final_calcium}
\end{equation}
}{
\begin{equation}
  p(t | \hat{r}) \propto {\cal H}(\hat{r} | f(t)).
\label{eq:final_calcium}
\end{equation}
}

We implement this method with a 1,000-element array uniformly sampled in log age from 1 Myr to 13,000 Myr, and evaluate Equation~\ref{eq:final_calcium} at each point in the array for the $\hat{r}$ of a single star.  These probabilities are then normalized to integrate to unity (accounting for uneven bin sizes), and provide the age posterior for that star.

\subsection{Lithium}
Overall, we follow the same procedure for lithium as calcium, with two major differences: lithium depletion has a strong $B-V$ color dependence (unlike the $R'_{HK}$ metric for calcium which was specifically formulated to be independent of color), and lithium measurements can have significant error bars or upper limits.

\subsubsection{Framework}

As with calcium, lithium equivalent width decreases with time, \deleted{but}with an astrophysical scatter about this trend for objects of the same age.  Following the framework we developed for calcium, we define functions for the mean equivalent width as a function of time ($i$), the standard deviation about that mean ($j$), and shape of the distribution function about the mean (\replaced{$k$}{${\cal K}$}).  These three functions are the lithium equivalents of $f$, $g$, and \replaced{$h$}{${\cal H}$} used above for calcium.  The mean $i(t,b)$ is decidedly a function of both age ($t$) and $B-V$ color ($b$).  However, when we consider the log of the equivalent width ($l$), the scatter about this mean appears to be independent of color, so we define $\replaced{k}{j}(t)$ as a function of time only.  The parameters of our model are $l$, $b$, and $t$, requiring a joint prior in all three for Bayes' equation, $p(l,b,t)$,

\replaced{
\begin{equation}
    p(l,b,t) = p(l|b,t) p(b) p(t) = k(l|i(t,b), j(t)) p(b) p(t)
\end{equation}
}
{
\begin{equation}
    p(l,b,t) = p(l|b,t) p(b) p(t) = {\cal K}(l|i(t,b), j(t)) p(b) p(t)
\end{equation}
}

\noindent where $p(b)$ is the prior on $B-V$ color, which we take to be flat, since we will generally have a precise measurement of color for a given star, and $p(t)$ is the prior on age, again flat for a constant star formation rate.

We assume a Gaussian likelihood for both $l$ and $b$, given measurements of $\hat{l}$ and $\hat{b}$, and measurement errors of $\sigma_{\hat{l}}$ and $\sigma_{\hat{b}}$, 

\begin{equation}
    \mathcal{L}(\hat{l},\hat{b}) = \mathcal{N}(10^{\hat{l}}|10^{l},\sigma_{\hat{l}}) \mathcal{N}(\hat{b}|b,\sigma_{\hat{b}})
\end{equation}

\noindent where 10 is raised to the power of $l$ and $\hat{l}$ since while $l$ is a log quantity, measurement errors are typically quoted in linear units (e.g. m\AA).   Combining likelihood and prior, and again assuming the evidence to be constant, we obtain an expression for the posterior

\replaced{
\begin{equation}
    p(l,b,t|\hat{l},\hat{b}) = \mathcal{N}(10^{\hat{l}}|10^{l},\sigma_{\hat{l}}) \mathcal{N}(\hat{b}|b,\sigma_{\hat{b}}) k(l|i(t,b), j(t))
\label{eq:lithium_posterior_almost_final}
\end{equation}
}
{
\begin{equation}
    p(l,b,t|\hat{l},\hat{b}) \propto \mathcal{N}(10^{\hat{l}}|10^{l},\sigma_{\hat{l}}) \mathcal{N}(\hat{b}|b,\sigma_{\hat{b}}) {\cal K}(l|i(t,b), j(t))
\label{eq:lithium_posterior_almost_final}
\end{equation}
}

\noindent which, when marginalized over $l$ and $b$, gives the marginalized posterior on age,

\begin{equation}
    p(t | \hat{l},\hat{b}) = \int \int p(l,b,t|\hat{l},\hat{b}) dl db.
\end{equation}

\noindent As with calcium, all that remains is to define the functions $i(t,b)$, $j(t)$, and $\replaced{k}{{\cal K}}(l | i(t,b), j(t))$ from our cluster data.

\subsubsection{The Color Dependence of Li EW}

\begin{figure*}
    \centering
    \includegraphics[page={1},width=\textwidth]{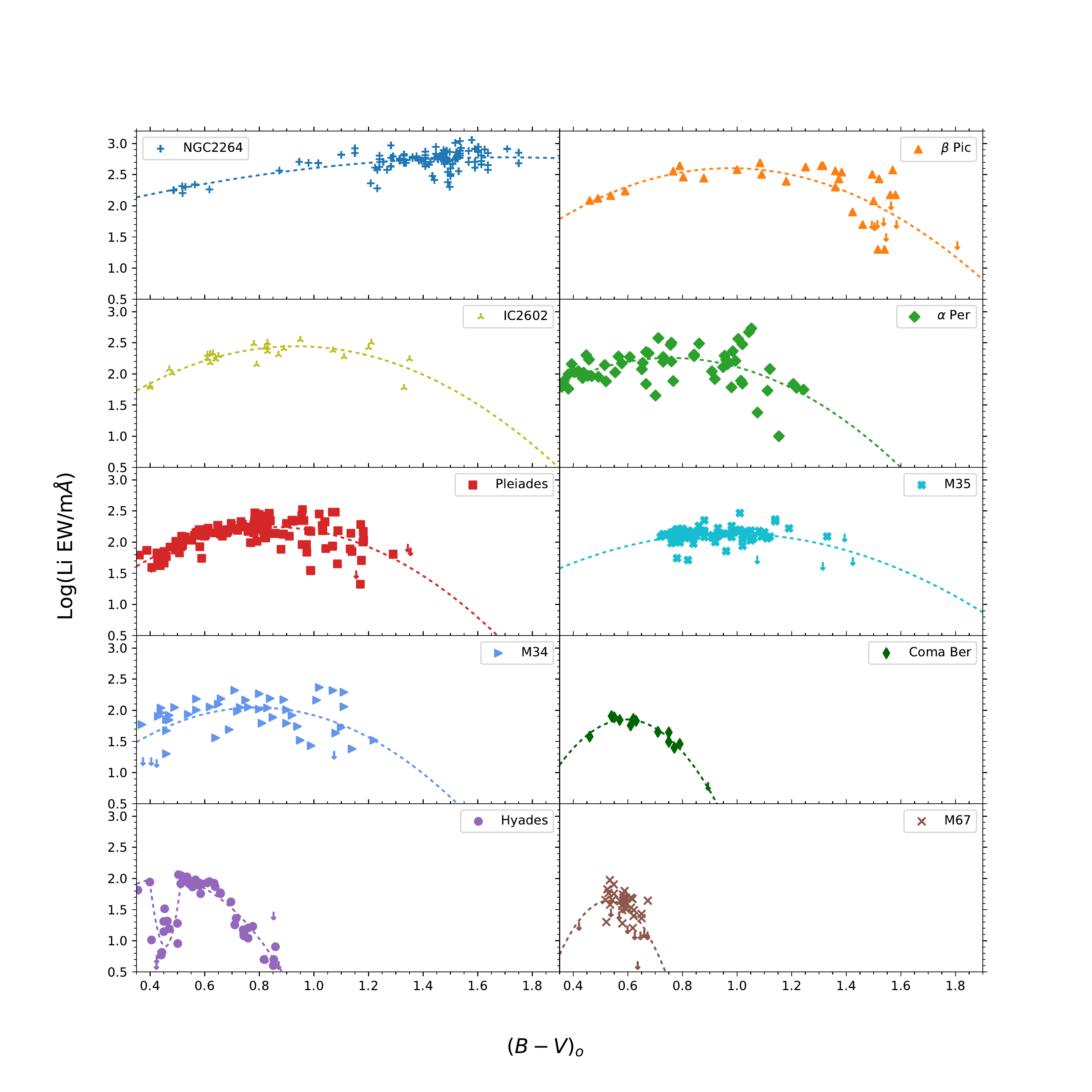}
    \caption{Lithium equivalent width measurements for our full dataset, and final fits to each cluster.  A second-order polynomial is a reasonable fit at all ages, outside of the lithium dip seen in the Hyades, which we model as a negative Gaussian. }
    \label{fig:li_vs_bv_all}
\end{figure*}

For a single cluster the log of the equivalent width, $l$, appears as a Gaussian or parabola as a function of $B-V$, as shown in Figure \ref{fig:li_vs_bv_all}.  The reddest stars and bluest stars in the cluster tend to have the smallest values for lithium EW while intermediate $B-V$ stars (G stars) have the largest lithium EW.  This behavior is the result of two primary processes. First, redder, lower-mass stars have deeper convective envelopes so more quickly convect lithium to deeper, hotter layers of the star where it is fused, resulting in faster depletion of lithium.  Meanwhile, blue stars have hotter photospheres so there are fewer lithium atoms in the ground state to absorb 6708 \AA\ light. Stars are expected to have uniform lithium abundance (N(Li)) at formation, but this translates to a range of Li EW values as a function of color, given the different photospheric temperatures across this range.  In addition to the Gaussian shape, the ‘lithium dip’ is observed for stars between $B-V$ of $\sim$0.36 and $\sim$0.42 (6900K and 6600K) for stars that are $\gtrsim$500 Myr, where there is a decrease in lithium abundance in this narrow range compared to stars on either side of the dip \citep{boesgaard:1986, balachandran:1995}. A suggested explanation for the lithium dip is that at the hot end of the dip, magnetic field strength is increasing with decreasing stellar mass, spinning down the outer layers of the star and creating turbulent mixing from internal shear between these layers and the faster-rotating core.  Then moving to the cooler end of the dip corresponds to the rise of internal gravity waves, which more efficiently spin down the core, so that there is less turbulent mixing \citep{talon:2010}. Under this model, surface lithium is preferentially destroyed in the narrow region of the lithium dip, while it is preserved on either side. 

For each cluster, we simultaneously fit both the mean\deleted{ of $l$, $i(t,b)$,} and the standard deviation\replaced{, $j(t)$,}{ of $l$} at a single value of $t = t_m$, where $t_m$ is the age of the given cluster $m$.  

\added{We take these fits, $i'(t_m,b)$ and $j'(t_m)$, as preliminary values for the mean, $i(t,b)$, and standard deviation, $j(t)$, evaluated at the age of the cluster.} 
We assume a functional form of a second-order polynomial for \replaced{$i(t_m,b)$}{$i'(t_m,b)$}, while at a single age the standard deviation\added{, $j'(t_m)$,} is a constant that does not depend on color\deleted{($j(t_m)$)}.  To fit these parameters we assumed a Gaussian likelihood, which for lithium detections takes the form

\replaced{
\begin{equation}
p(l|t_m,b) = \mathcal{N}(\hat{l}|l,\sigma_{\hat{l}}) =  \frac{1}{\sqrt{2\pi}j(t_m)} e^{-\frac{(\hat{l} - i(t_m,b))^2}{2j(t_m)^2}}
\label{eq:likelihood}
\end{equation}
}{
\begin{equation}
p(\hat{l}|t_m,b) = \mathcal{N}(\hat{l}|l,\sigma_{\hat{l}}) =  \frac{1}{\sqrt{2\pi}j'(t_m)} e^{-\frac{(\hat{l} - i'(t_m,b))^2}{2j'(t_m)^2}} .
\label{eq:likelihood}
\end{equation}
}

\noindent For lithium upper limits ($\hat{u}$) we represent the likelihood as the integral of the Gaussian function from -$\infty$ to the upper limit $\hat{u}$,  

\replaced{
\begin{equation}
p(u|t_m,b) = \int_{-\infty}^{u} \frac{1}{\sqrt{2\pi}j(t_m)}e^{-\frac{(l - i(t_m,b))^2}{2j(t_m)^2}} dl.
\label{eq:ul_likelihood}
\end{equation}
}{
\begin{equation}
p(\hat{u}|t_m,b) = \int_{-\infty}^{\hat{u}} \frac{1}{\sqrt{2\pi}j'(t_m)}e^{-\frac{(l - i'(t_m,b))^2}{2j'(t_m)^2}} dl,
\label{eq:ul_likelihood}
\end{equation}
}
\noindent and then fit these four parameters (three for the polynomial in $b$ that defines \replaced{$i(t_m,b)$}{$i'(t_m,b)$}, and one for the standard deviation, \replaced{$j(t_m)$}{$j'(t_m)$}.  The fit itself is performed by assigning one of these likelihoods to each star, based on whether there is a lithium measurement or upper limit, then maximizing the product of  likelihoods over all cluster stars.

\begin{figure*}
  \centering
  \begin{tabular}{c @{\qquad} c }
    \includegraphics[page={1},width=.48\textwidth]{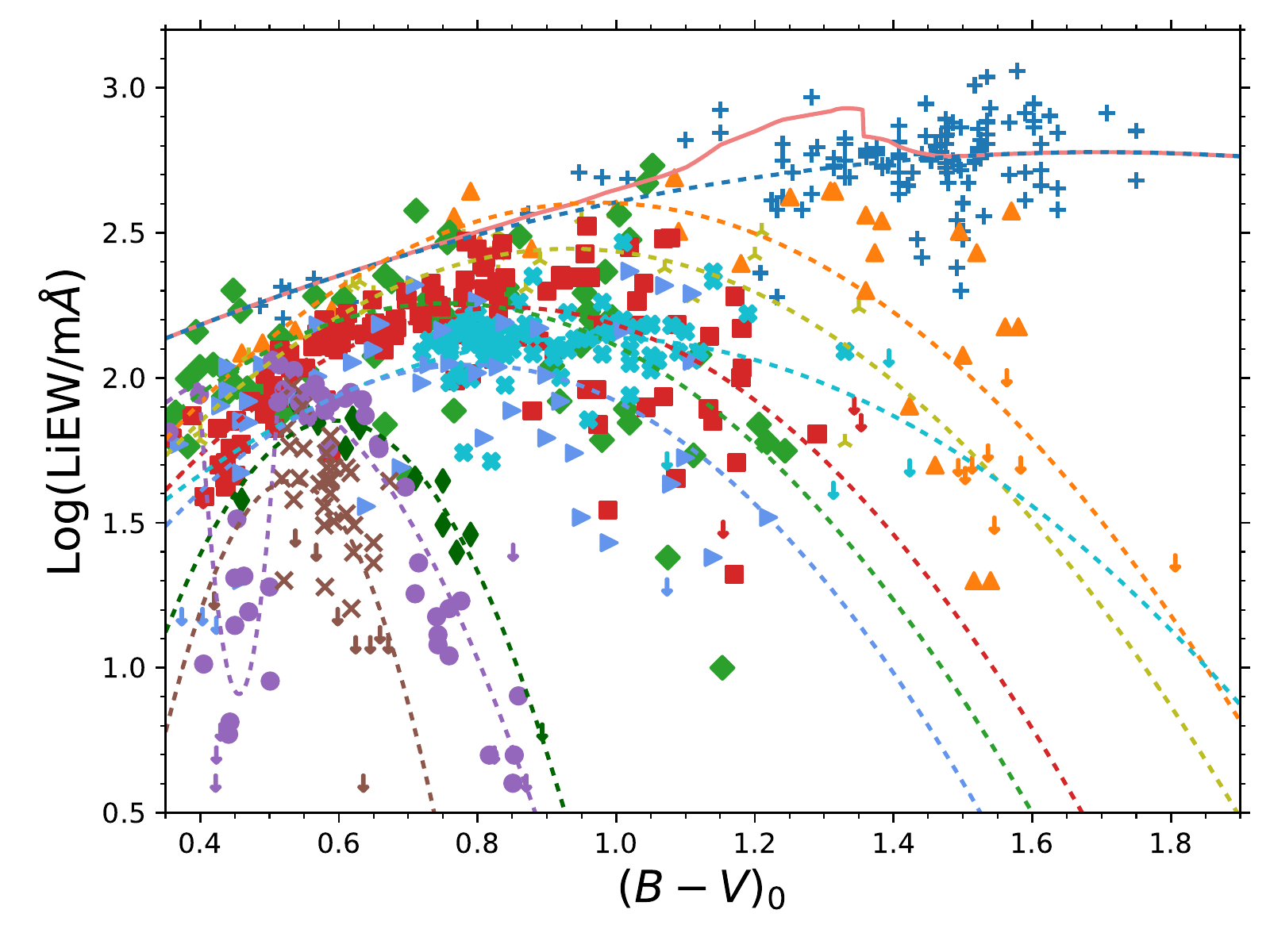} &
    \includegraphics[page={2},width=.48\textwidth]{lithium_metal_vs_bv.pdf} \\
    \small (a) Full lithium dataset and fits to each cluster & \small (b) Fits only
  \end{tabular}
  \caption{Lithium equivalent width measurements for our full dataset, and final fits to each cluster. As expected, lithium equivalent width decreases monotonically over time, but as a strong function of $B-V$ color. Primordial Li EW is estimated from MIST isochrones in conjunction with our fit to NGC2264.}
\label{fig:li_vs_bv}
\end{figure*}

The lithium dip is clear in the $\sim$700 Myr Hyades dataset, so we fit an inverted Gaussian to the dip ($0.39 < B-V < 0.52$), and a second-order polynomial to the stars outside the dip.  There is no clear evidence for a lithium dip at younger ages in M34 ($\sim$200 Myr), and by $\sim$4 Gyr, stars have evolved off the main sequence leaving no stars bluer than $B-V \approx 0.5$ in M67. Fits to each cluster are shown in Figures~\ref{fig:li_vs_bv_all} and \ref{fig:li_vs_bv}.

\subsubsection{Li EW as a function of age}
Unlike $R'_{HK}$, Li EW varies substantially across both age and $B-V$. As a result, the decline in lithium equivalent width as a function of age must be fit across multiple slices of $B-V$. The polynomials fit to each cluster \added{(Fig \ref{fig:li_vs_bv_all})} define \replaced{$i(t_m,b)$}{our preliminary fits $i'(t_m,b)$} as a function of $B-V$ at the age for each individual cluster ($t_m$). We examine 64 $B-V$ slices uniformly spaced between $B-V$ of 0.35 and 1.9. At each slice $n$, then, we have 10 values of \replaced{$i(t_m,b_n)$}{$i'(t_m,b)$} corresponding to our 10 cluster datasets to which we add two additional points from primordial Li EW and Blue Lithium Depletion Boundary (described below), and from these we determine the 64 fits \replaced{to $i(t_m,b_n)$}{$i(t,b_n)$} as a function of age.

To help constrain the young end of the fits of Li EW and age, we approximate primordial Li EW using the MIST model isochrones \citep{MIST2016} in conjunction with our NGC2264 fit.  In particular, we seek to extend the fit to this $\sim$5 Myr cluster to the first age point in our grid, 1 Myr. At every $B-V$ value we determine the corresponding effective temperature using the conversions from \citet{pecaut:2013}; then we find the Li abundance, N(Li), and initial stellar mass at 5 Myr from the MIST isochrones. We then find the Li abundance from the same initial mass star using the 1 Myr isochrones.  $T_{eff}$ and Li abundance are converted to Li equivalent width using the curve of growth in \citet{soderblom:1993pleiades} for $T_{eff} > 4000 K$ and \citet{zapatero_osorio2002} for $T_{eff} \leq 4000 K$.  The difference in Li EW between 1 Myr and 5 Myr is added to the fit to NGC2264 to determine the primordial Li EW at every $B-V$ value (Figure \ref{fig:li_vs_bv}).  \replaced{The MIST isochrone correction}{The found change in Li EW} between 1 and 5 Myr is only significant between $0.8 \lesssim B-V \lesssim 1.4$, and is negligible elsewhere.

\begin{figure}[htp]
\centering
\includegraphics[width=\columnwidth]{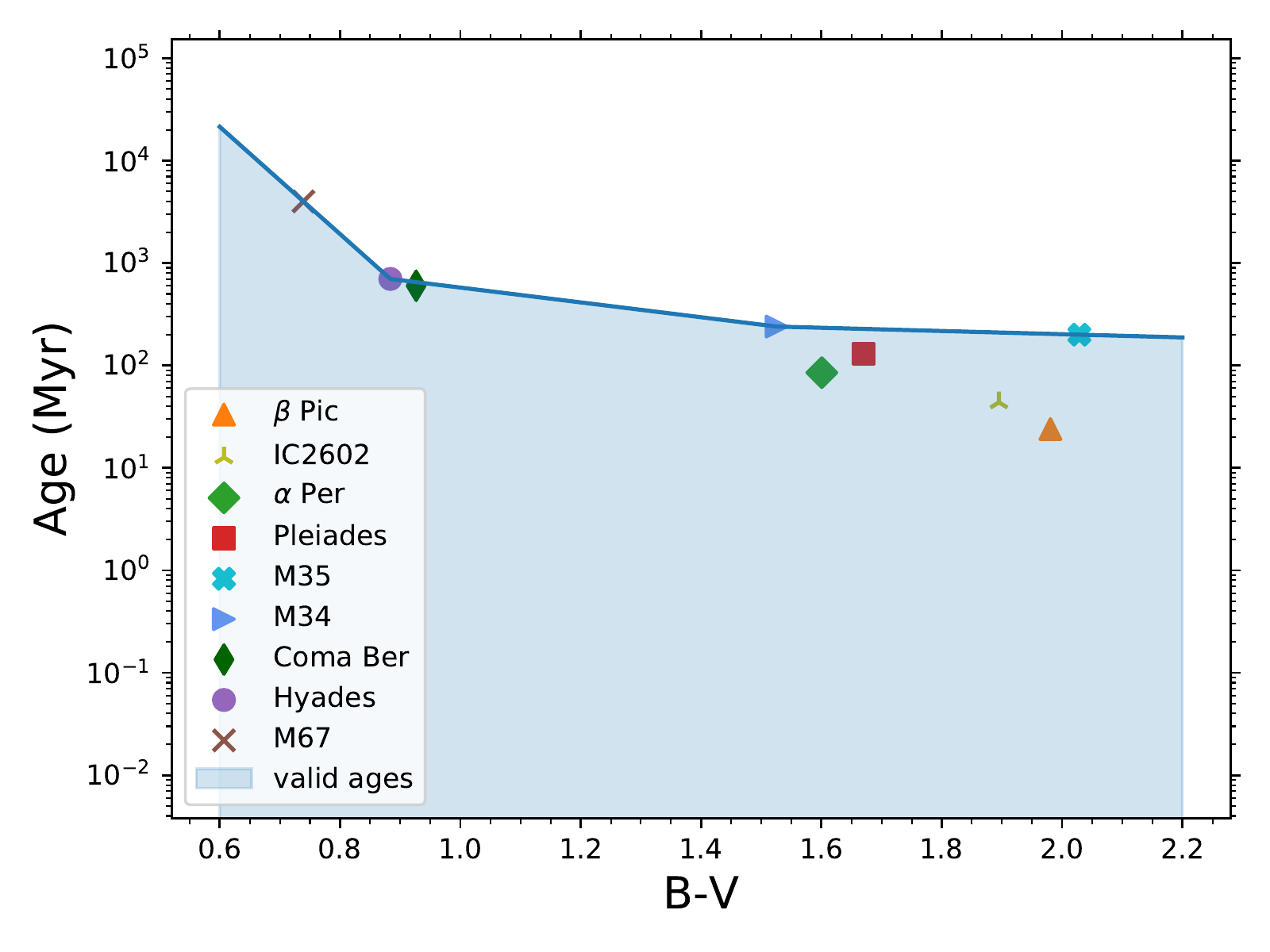}
\caption{We introduce the concept of the Blue Lithium Depletion Boundary (BLDB) -- which represents the age at each $B-V$ slice where lithium equivalent width drops below 3.2 mA -- to constrain the lithium abundance at the oldest ages.  Each point represents the $B-V$ magnitude where our polynomial fit to each cluster (\replaced{log(Li EW) vs $B-V$}{$i'(t_m,b)$}) goes below log(Li EW) = 0.5 or 3.2 m\AA, which we adopt as the lowest detectable equivalent width of the lithium line.  Redward of the BLDB point, we expect all stars in the cluster have no detectable lithium 6708 \AA\ absorption.  We adopt a piecewise linear fit such that all clusters are at or below the fit.  }
\label{fig:blbd}
\end{figure}

We define the Blue Lithium Depletion Boundary (BLDB) as the $B-V$ color for which stars redder than this boundary have no detectable lithium absorption, which we use to help constrain the older and redder range of fits to Li EW against age.  Since the redder stars have deeper convective envelopes they burn lithium faster than the bluer stars in the cluster.  As a result the nested polynomials of Figure \ref{fig:li_vs_bv} generally get narrower and move blueward over time, and thus the BLDB point moves blueward with increasing age.   The BLDB is distinct from the classical lithium depletion boundary (LDB) which moves redder over time as a cluster’s high-mass brown dwarfs deplete their lithium, while at the same time all brown dwarfs evolve to redder colors as they cool over time, outside the brief deuterium burning phase.  We have defined the BLDB in order to add an additional data point to our fits for $B-V$ $>$ 0.7, most important for constraining the ages of stars with $B-V$ $>$ 1.4 for which there are fewer literature measurements, especially at older ages.  For each $B-V$ slice redward of 0.7, the fit to BLDB points gives an approximation of maximum age associated with log(Li EW) = 0.5, or Li EW = 3.2 m\AA\ (Figure \ref{fig:blbd}).

\begin{figure*}
  \centering
    \includegraphics[width=\textwidth]{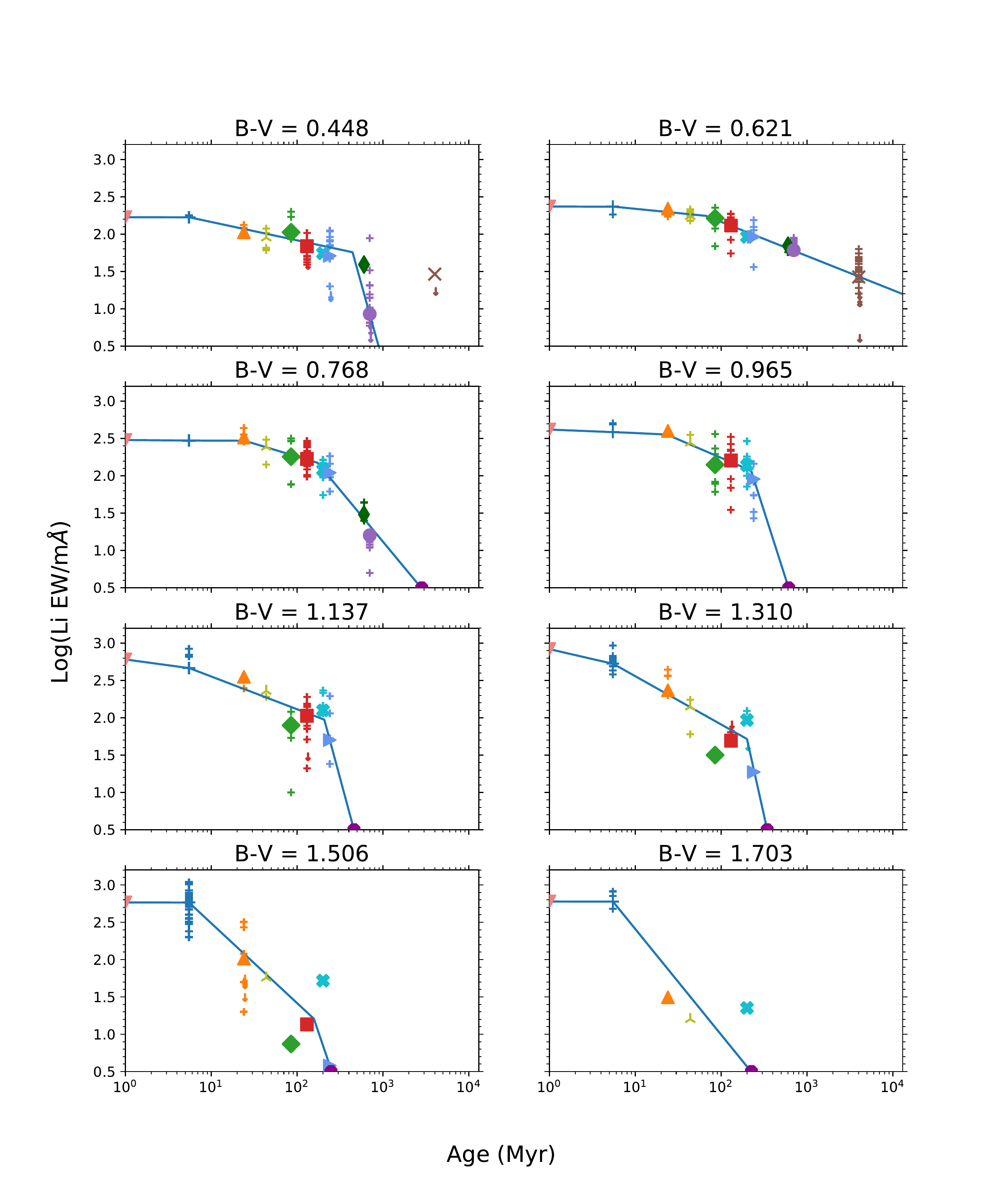} 
\caption{Examples of Li EW fits as a function of age $i(t,b_n)$, for eight out of 64 $B-V$ slices between 0.35 and 1.9 mags. Stars from each cluster within 0.05 mags of the $B-V$ slice are shown as small crosses if detections or downward-facing arrows if upper limits. Cluster symbols are as in Figure \ref{fig:li_vs_bv}, with an additional magenta BLDB point at log(Li EW)=0.5. The cluster means were fit with a flexible piecewise-linear function fixed to the primordial lithium point and NGC2264.  Cluster means were weighted to give those with the most stars at each $B-V$ slice the most weight. }
\label{fig:li_vs_age}
\end{figure*}

For each value of $B-V$ we use the mean value of $l$, \replaced{$i(t_m,b_n)$}{$i'(t_m,b_n)$}, from each cluster in addition to the primordial lithium point and BLDB point to fit the intermediate ages between the cluster ages and complete our grid.  Unlike calcium, where the fits to individual clusters were independent of $B-V$, for lithium there is a strong $B-V$ dependence, and for redder regions, the fit to the mean equivalent width reaches unphysically small values. When cluster means drop below $\sim$3m\AA\ (0.5 on the log scale), we don’t expect any detections, and clusters with \replaced{$i(t_m,b)$}{$i'(t_m,b)$} below this value are not included in the fitting process.  

As in Figure \ref{fig:li_vs_age}, we fit a  2-4 segment piecewise function to the cluster means, primordial Li EW, and BLDB point.  The first segment is always between the Primordial Li EW value and NGC2264 and the fit is constrained to decrease monotonically with age.  Additionally, for $B-V$ slices inside the lithium dip (0.41 $\le$ $B-V$ $\le$ 0.51), the final piecewise segment is constrained to go through the Hyades point.  The locations of the segment breaks (except for the first break at NGC2264) were free parameters.  Weights for the cluster means were determined based on the relative proportion of stars the cluster had at a given $B-V$ slice in relation to the total number of stars.  The BLDB point is given uncertainty about 0.15 dex compared to 1 dex for clusters in not well constrained regions. Although for some $B-V$ ranges different functional forms were good fits to the decrease in lithium over time, only the piecewise function was flexible enough to capture the shape more generally.

\subsubsection{Astrophysical Scatter}
 
\begin{figure*}
\centering
    \begin{tabular}{c @{\qquad} c }
    \includegraphics[page={1},width=.48\textwidth]{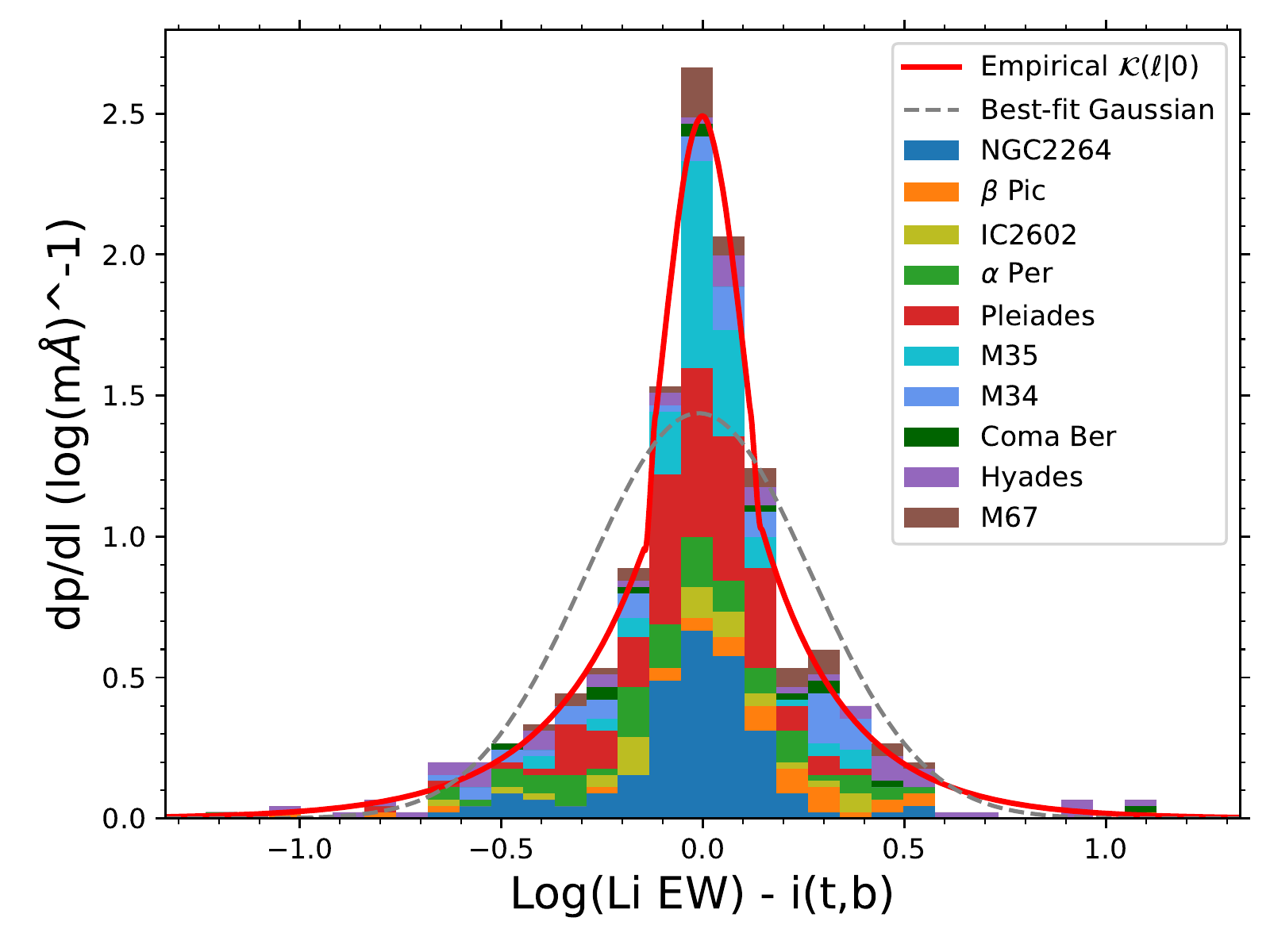} &
    \includegraphics[page={2},width=.48\textwidth]{lithium_hist_fit.pdf} \\
    \end{tabular}
\caption{(left) The numerically determined function \replaced{$k(l|i(t,b)$ is the red line, plotted against the }{${\cal K}(l|i(t,b))$, which defines the astrophysical scatter in lithium abundance for stars of the same age and color, is plotted as a red line, along with the} residuals of all cluster stars to the fit $i(t,b)$, and (right) the corresponding CDF.  Similar to calcium, the numerical PDF has exponentially decreasing tails similar to a Gaussian distribution but is significantly more peaked.}
\label{fig:lithium_pdf}
\end{figure*}

With our grid of 64 $B-V$ slices and 1000 age slices for mean equivalent width of lithium, $i(t,b)$, we next empirically determine the distribution of the residuals, $\replaced{k}{{\cal K}}(l|i(t,b),j(t))$, as we did with calcium (Figure \ref{fig:lithium_pdf}).  Residuals are with respect to the value of $i(t,b)$ evaluated at the age of each cluster \added{and the $B-V$ value of the star}, and upper limits are not considered in this step. As with calcium, we smooth the CDF of the residuals with Savitzky–Golay filters and take the derivative to convert to a PDF. Then we fit exponential functions to the two tails, which we connect with the smoothed PDF, then normalize to have integral unity, defining $\replaced{k}{{\cal K}}(l|i(t,b),j(t))$. \added{We also center the distribution at zero by subtracting off the residual median value of 0.033, which ensures that ${\cal K}$ does not introduce a systematic bias toward older ages.}  Unlike calcium, we find \replaced{no evidence that $j(t)$, the standard deviation of the residuals, varies with time.}{no evidence for even a weak dependence on time of the standard deviation of the residuals ($j(t)$).}  As a result, the shape of \replaced{$k$}{${\cal K}$} is not a function of age or color, while the mean value is.

\subsubsection{Lithium posterior}

Since we see no evidence for an age dependence in the astrophysical scatter, we take $j(t)$ to be a constant, and slightly rewrite our posterior from Equation~\ref{eq:lithium_posterior_almost_final},

\replaced{
\begin{equation}
    p(l,b,t|\hat{l},\hat{b}) = \mathcal{N}(10^{\hat{l}}|10^{l},\sigma_{\hat{l}}) \mathcal{N}(\hat{b}|b,\sigma_{\hat{b}})k(l|i(t,b)).
\label{eq:lithium_posterior_final}
\end{equation}
}{
\begin{equation}
    p(l,b,t|\hat{l},\hat{b}) \propto \mathcal{N}(10^{\hat{l}}|10^{l},\sigma_{\hat{l}}) \mathcal{N}(\hat{b}|b,\sigma_{\hat{b}}) {\cal K}(l|i(t,b)).
\label{eq:lithium_posterior_final}
\end{equation}
}
To determine the age posterior for a single star, we construct a dense grid covering $B-V$ from 0.35 to 1.9 and age from 1-13,000 Myr. We use a grid of 64,000 elements (64 $B-V$ x 1000 age, logarithmically spaced in age), with the mean lithium abundance $i(t,b)$ calculated at each grid point.  At each combination of $(t,b)$, we first marginalize over $l$ by multiplying $\replaced{k}{{\cal K}}(l|i(t,b))$ (our prior) by $\mathcal{N}(10^{\hat{l}}|10^{l},\sigma_{\hat{l}})$, a Gaussian representing the measurement of lithium equivalent width and the associated measurement error, over an array of 1000 elements, logarithmically spaced in $l$ between 0.5 and 1585m\AA.  To do this multiplication, however, we first convert $\replaced{k}{{\cal K}}(l|i(t,b))$ to a function (similar to a log-normal) in linear space as the likelihood for $\hat{l}$, $\mathcal{N}(10^{\hat{l}}|10^{l},\sigma_{\hat{l}})$, is defined in linear space. If no measurement uncertainty is given we use a default error for $\sigma_{\hat{l}}$ of 15m\AA, which is noted as a typical error by \citet{soderblom:1993pleiades}.  The products of these functions evaluated at all 1000 points are then summed, which gives the probability at that specific $(t,b)$ gridpoint. We then marginalize over $B-V$ color by weighting each (t,b) gridpoint by the Gaussian likelihood for $b$, $\mathcal{N}(\hat{b}|b,\sigma_{\hat{b}})$, (assuming $\sigma_{\hat{b}}$ = 0.01 mags if no error is given) and summing over the product.
To minimize computation time, instead of computing this probability at all (t,b) locations, we only evaluate gridpoints at 15 sampled values of $b$ within 4$\cdot \sigma_{\hat{b}}$ of $\hat{b}$, with all others set to 0.

Having marginalized over both $l$ and $b$, we are left with a marginalized posterior over only age, $p(t|\hat{l},\hat{b})$.

If the measurement $\hat{l}$ is an upper limit $\hat{u}$, we instead integrate $\replaced{k}{{\cal K}}(l|i(t,b))$ from $-\infty$ to $\hat{u}$ to find the probability at each (t,b) gridpoint.  Thus upper limits result in a plateau of probability at old ages, with a rapid drop-off toward younger ages.

\section{Validation}

\subsection{Self-consistency of age posteriors}
\begin{figure*}
  \centering
  \includegraphics[width=1.02\textwidth]{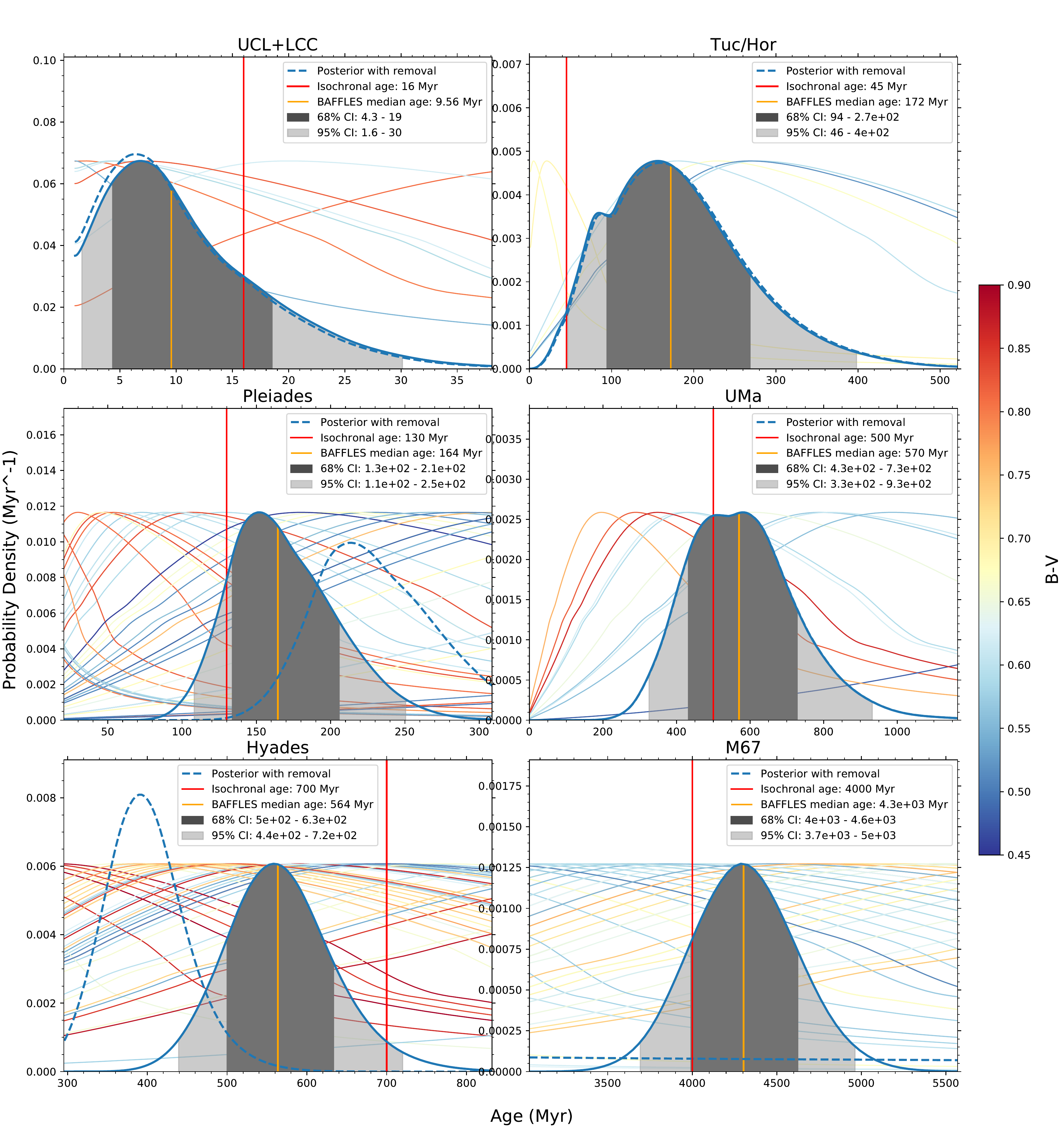}
\caption{We test the validity of our calcium age posteriors by considering the product of PDFs from every star in one of our benchmark clusters, which should represent the PDF of the cluster age.  We calculate the age posteriors for each star in the cluster (shown color-coded by $B-V$ and scaled to common height in the background), and finally multiply the age posteriors together to get the posterior product. The blue dashed line is the posterior product produced if we first omit the cluster from those used to calibrate {\tt BAFFLES}. ``Isochronal Age" represents the more robustly determined ages from Table \ref{table:clusters} that we use as the ages of our benchmark clusters. The mostly 1-2 $\sigma$ agreement suggests the median age and uncertainties we find with {\tt BAFFLES} are reasonable, though uncertainties reported by {\tt BAFFLES} may be \replaced{somewhat}{slightly} underestimated.}
\label{fig:ca_posterior_product}
\end{figure*}

\begin{figure*}
  \centering
  \includegraphics[width=1.02\textwidth]{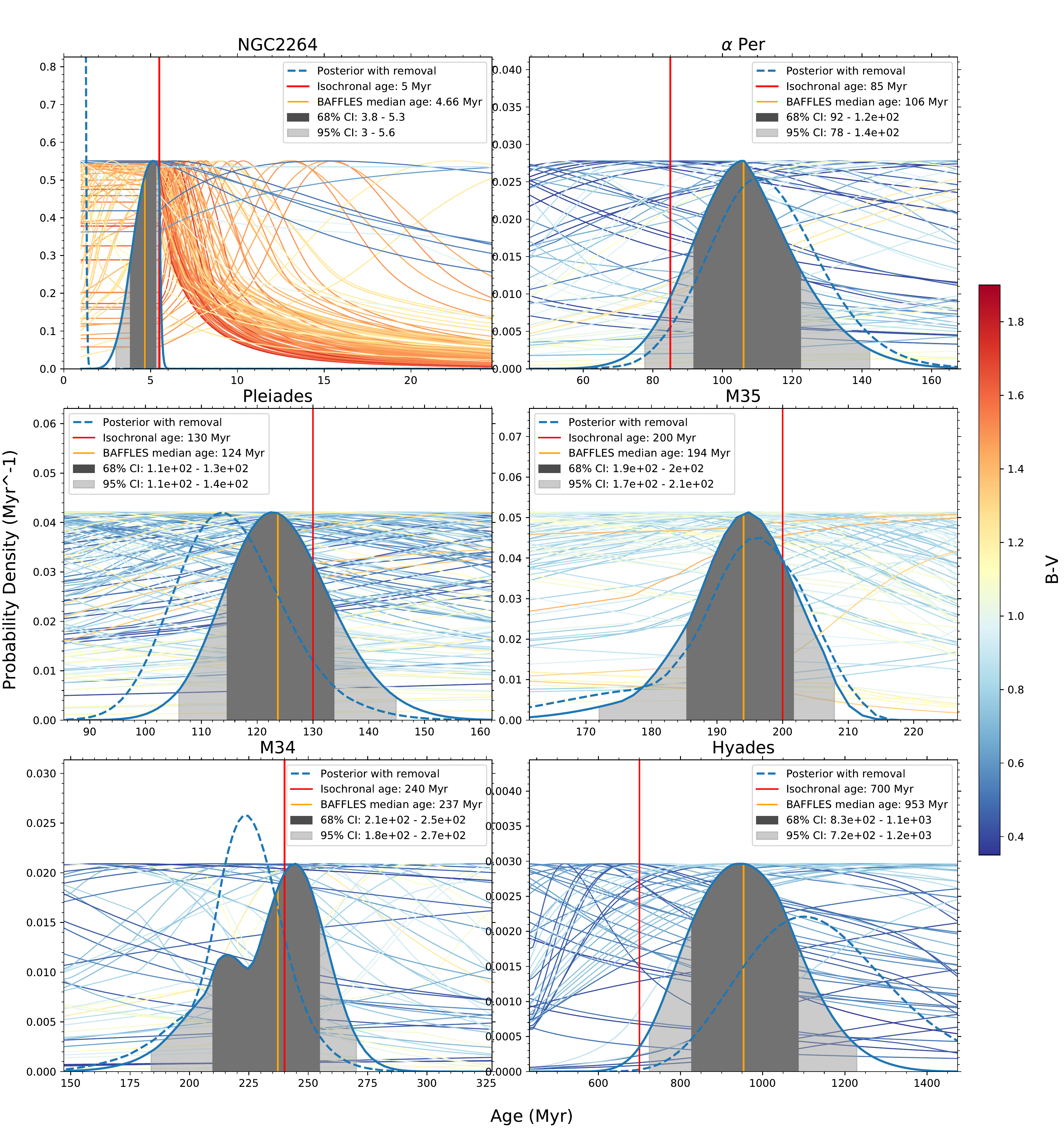}
\caption{Following Figure \ref{fig:ca_posterior_product}, we test the validity of our lithium posteriors. Our ages are consistent with isochronal ages to within 68\% confidence intervals for \replaced{7 of our 10 clusters, with $\alpha$ Per, Hyades, and M67 at the 98.8\%, 99.9\%, and 97.4\% confidence interval, respectively.}{6 of our 10 clusters, and to within 95\% confidence intervals for 9/10 clusters.} \deleted{If we recompute the Hyades age with only stars of $B-V > 0.55$, to exclude the `lithium dip', the isochronal age of 700 Myr falls within our 75\% confidence interval, which suggests the Hyades age discrepancy is due to how we handle the dip.}}
\label{fig:li_posterior_product}
\end{figure*}

To test {\tt BAFFLES} for self-consistency, we compare the posteriors for stars in moving groups and associations to the known ages of the groups, which we show for some clusters in Figures \ref{fig:ca_posterior_product} and \ref{fig:li_posterior_product}. We compute posteriors for each star in a cluster and then multiply the posteriors together, assuming the age determination for each star is independent, to produce a probability density function for the age of the cluster as a whole.  As an additional test, we repeat the process, but beforehand remove the target cluster from the input clusters used to fit \replaced{mean $R'_{HK}$ vs Age}{$f(t)$ and $i(t,b)$} (though leave the cluster in for \replaced{scatter vs age fits and computing the PDF of astrophysical scatter}{computing ${\cal H}(r|f(t))$ and ${\cal K}(l|i(t,b))$}).

\replaced{
This validation indicates that while there is no systematic offset in derived {\tt BAFFLES} ages, there is evidence that uncertainties are underestimated. When we compare the isochronal age of each cluster to the posterior product, we find no trend in the direction of the offset between the median of the posterior and the isochronal age.  The degree of offset from the median, however, suggests uncertainties may be underestimated.  For the nine calcium clusters, UCL+LCC, $\beta$ Pic, Tuc/Hor, $\alpha$ Per, Pleiades, UMa, Hyades, and M67, we find the isochronal age is within the 12.1\%, 91.7\%, 98.5\%, 99.5\%, 6.5\%, 23.9\%, 36.1\%, 99.0\%, and 89.4\% confidence interval, respectively.  We would expect two thirds of the clusters to fall within the 68\% confidence interval, and most to fall within 95\%, but that is not the case here. Though there is limited evidence given the small number of clusters, the largest outliers are at $\sim$20 Myr and $\gtrsim$700 Myr, suggesting issues either with the fit to the mean ($f$) or fit to the scatter ($g$). Three of the five most discrepant clusters, UCL+LCC, $\beta$ Pic, and Tuc/Hor have 8, 6, and 6 stars in our dataset, respectively.  As a result, both the median and scatter of these clusters is poorly determined.  From Figure \ref{fig:ca_vs_age}, UCL+LCC are slightly above the fit, and Tuc/Hor is slightly below.  Also, the standard deviation we measure from those three clusters in Figure~\ref{fig:calcium_scatter_vs_age} vary greatly from 0.7, to 0.21, to 0.13 dex.  Going forward, larger sample sizes at these young ages are needed to better determine the time evolution of $R'_{HK}$.  For now we advise caution that {\tt BAFFLES} posteriors may somewhat underestimate the errors from calcium, especially in less-well-sampled age regimes.
This validation indicates that while there is no systematic offset in derived {\tt BAFFLES} ages, there is evidence that uncertainties are underestimated.

}{

{\tt BAFFLES} ages determined from calcium posterior products matched well with isochronal ages (Figure \ref{fig:ca_posterior_product}), though have a slight shift toward older ages. We find 6/9 clusters have ages older than their isochronal age, with only UCL+LCC, $\alpha$ Per, and Hyades being younger. Unsurprisingly, all three clusters lie above the fit in Figure \ref{fig:ca_vs_age}. For the nine calcium clusters, Upper Sco, UCL+LCC, $\beta$ Pic, Tuc/Hor, $\alpha$ Per, Pleiades, UMa, Hyades, and M67, we find the isochronal age is within the 3.42\%, 54.7\%, 75.2\%, 95.8\%, 88.1\%, 74.5\%, 35.7\%, 92.6\%, and 66.9\% confidence interval, respectively.  We would expect two thirds of the clusters to fall within the 68\% confidence interval, and nearly all to fall within 95\%, but here we have 4/9 within the 68\% CI and 8/9 within the 95\% CI. Notably, the largest outlier is Tuc/Hor (isochronal age within 95.8\% CI) for which there are only 6 calcium stars.  From Figure \ref{fig:ca_vs_age}, Tuc/Hor appears to be the cluster farthest from the fit, and its distance below the fit pushes the predicted ages of its stars older. Going forward, larger sample sizes at these young ages are needed to better determine the time evolution of $R'_{HK}$.  For now we advise slight caution that {\tt BAFFLES} posteriors may \added{slightly} underestimate the errors from calcium, especially in less-well-sampled age regimes.
}

A similar posterior product check with lithium clusters found \added{good} agreement with isochronal values (as seen in Figure \ref{fig:li_posterior_product}). \replaced{NGC2264, $\beta$ Pic, IC2602, Pleiades, M35, M34, and Coma Ber have isochronal ages within the 68.27\% confidence interval.}{We find 6/10 clusters have isochronal ages within the 68\% confidence interval: $\beta$ Pic, IC2602, Pleiades, M35, M34, and Coma Ber.} \added{NGC2264,} $\alpha$ Per, Hyades and M67 have isochronal ages\deleted{that are younger than the median {\tt BAFFLES} ages, and are} within the \replaced{98.8, 99.9 and 97.4\%}{90.1\%, 85.9\%, 97.7\% and 85.2\%} confidence intervals, respectively\added{, making 9/10 clusters within the 95\% CI}. \added{We also find that half the clusters are younger than their isochronal ages (NGC2264, IC2602, Pleiades, M35, M34), while the other half are older, indicating no systematic offset in ages.}  The offset in the Hyades is \replaced{likely due to how the lithium dip is handled, with an older age assigned to dip stars compared to the other stars in the cluster.}{likely due to 3 upper limits in the lithium dip with log(Li EW) between $\sim 0.6-0.8$ dex, which significantly pull the posterior product to older ages.} Computing the age of the Hyades \replaced{with only stars of $B-V > 0.55$ (excluding all dip stars)}{while excluding these 3 upper limits (leaving 44 detections and 3 other upper limits)}, \added{{\tt BAFFLES} reports an age of 798 Myr, and} the isochronal age falls within our \replaced{75}{74}\% confidence interval. \added{Thus, it is likely more work needs to be done to properly model the lithium dip to produce robust posteriors for stars within the dip}. As with calcium, a product of posteriors is very sensitive to each individual posterior, so that a single non-member, or errors in color or lithium abundance, can move the product significantly from the age of the group as a whole. We conclude that the age posteriors generated by {\tt BAFFLES} from lithium abundances are consistent with the ages of our benchmark clusters.

\subsection{Moving Groups}

\begin{figure*}
  \centering
  \begin{tabular}{c @{\qquad} c }
    \includegraphics[page={1},width=.48\textwidth]{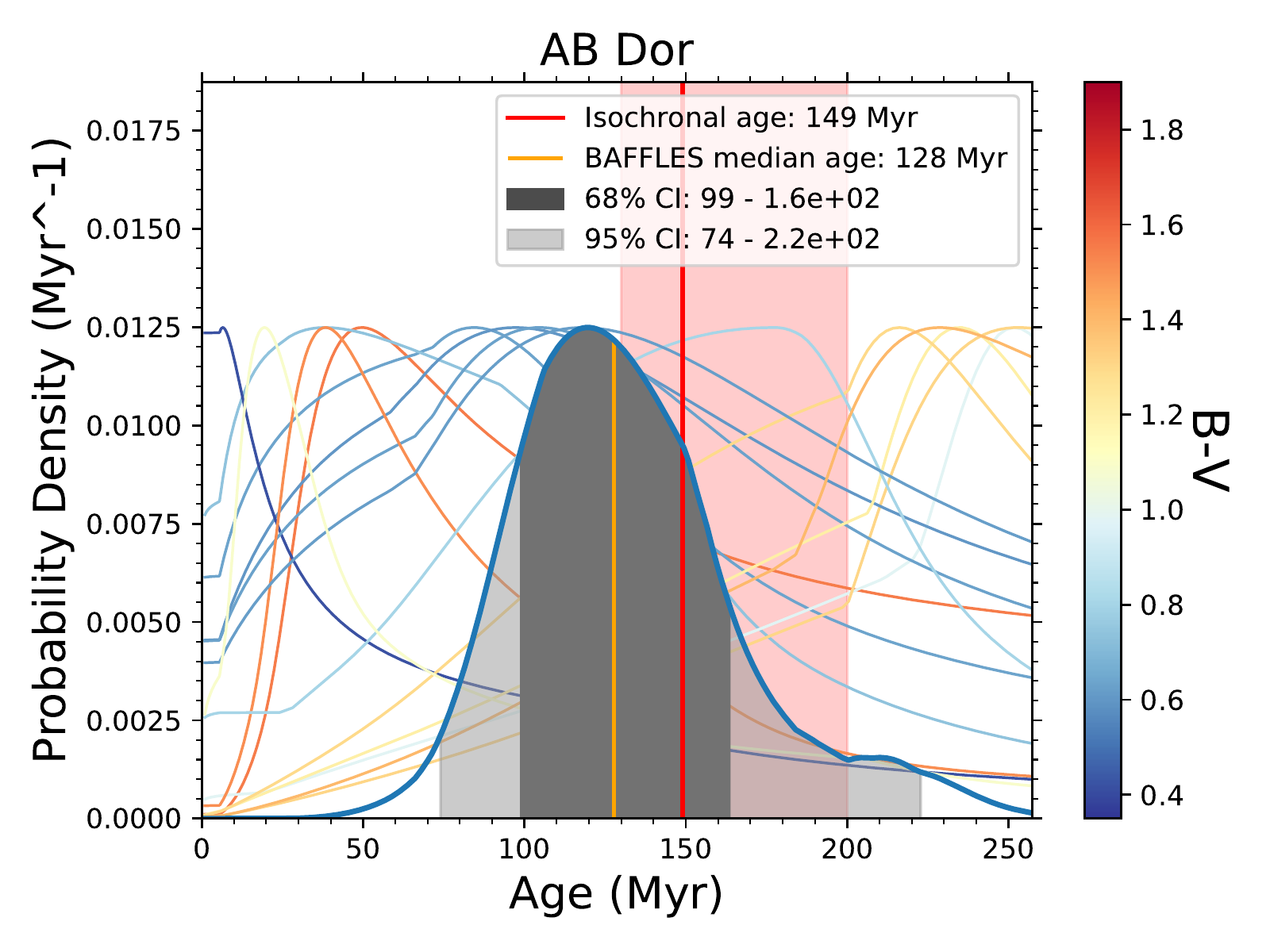} &
    \includegraphics[page={2},width=.48\textwidth]{moving_group_age.pdf} \\
  \end{tabular}
\caption{We compute age posteriors for AB Dor and Tuc/Hor from the product of the posteriors of stars in each moving group. Our computed ages agree with isochronal values to within 1$\sigma$, suggesting that our lithium-derived posteriors are generally accurate. Data for AB Dor and Tuc/Hor is from \citet{mentuch:2008} and isochronal ages are from \citet{bell:2015}.}
\label{fig:moving_groups}
\end{figure*}

We further examine the accuracy of {\tt BAFFLES} age posteriors by considering the ages derived for multiple stars in moving groups not included in our set of benchmark clusters.  As before, we compute age PDFs for each star in the moving group and then multiply the PDFs together to find an age for the group as a whole, which we compare to isochronal ages from \citet{bell:2015} (Figure \ref{fig:moving_groups}).     

Lithium equivalent widths for AB Dor and Tuc/Hor are from \citet{mentuch:2008}, and $B-V$ magnitudes \replaced{are from the Simbad online services.}{used are provided in Table \ref{table:B_V_ref} in the appendix}. From AB Dor we removed the binaries HD 13482A, HD 13482B, HD 17332B, HD 217379N, HD 217379S and also stars with large $B-V$ uncertainties (error $\gtrsim$0.15 mags): BD+21 418B. From Tuc/Hor we removed the binaries: AF Hor, BS Ind, HIP 116748N, HIP 116749S, TYC 7065-0879N, TYC 7065-0879S as well as stars with error $\gtrsim$0.15 mags: EXO 0235.2-5216, CD-58 553, Smethells 86, CT Tuc, Smethells 165, Smethells 173.

We derive ages for AB Dor: \replaced{$135_{-26}^{+30}$ Myr}{$127_{-28}^{+35}$ Myr}, and Tuc/Hor: \replaced{$42_{-8}^{+10}$ Myr}{$35_{-10}^{+11}$ Myr}, as in Figure \ref{fig:moving_groups}.  These ages are within 1$\sigma$ of isochronal ages -- AB Dor $149_{-19}^{+51}$, Tuc/Hor $45_{-4}^{+4}$ Myr -- from \citet{bell:2015}. We caution against using the ages we derive for these moving groups, however, since our posterior products can be significantly biased by a single star with incorrect values (either lithium abundance, $B$ or $V$), or with an incorrect membership determination.

\section{Analysis}

\subsection{Notable Stars: TW PsA, HR 2562, and HD 206893}
We show examples of ages derived using {\tt BAFFLES} for three field stars associated with substellar companions: the brown dwarf hosts HR 2562 and HD 206893, and the stellar companion to the exoplanet host Fomalhaut, TW PsA. Age sets the formation timescale for these substellar companions, and in the case of the brown dwarfs, the model mass derived for these objects depends directly on the assumed age.

\begin{figure*}
  \centering
  \begin{tabular}{c @{\qquad} c }
    \includegraphics[page={1},width=.48\textwidth]{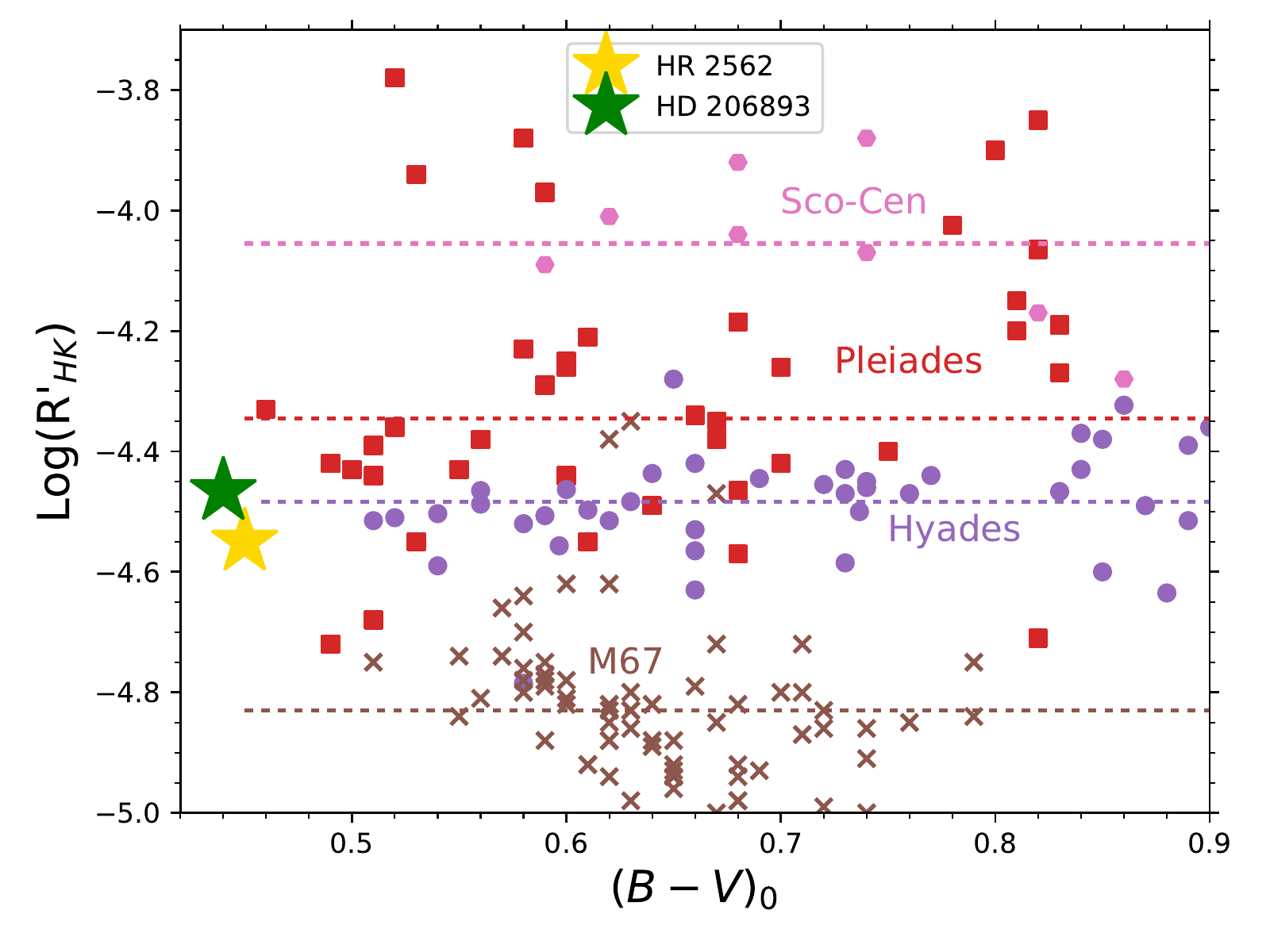} &
    \includegraphics[page={2},width=.48\textwidth]{notable_stars.pdf} \\
    \includegraphics[page={5},width=.48\textwidth]{notable_stars.pdf} &
    \includegraphics[page={6},width=.48\textwidth]{notable_stars.pdf} \\
    \includegraphics[page={3},width=.48\textwidth]{notable_stars.pdf} &
    \includegraphics[page={4},width=.48\textwidth]{notable_stars.pdf} \\
  \end{tabular}
\caption{{\tt BAFFLES} age posteriors for three notable field stars from calcium $R'_{HK}$ and lithium equivalent width.  The top panel shows the measurements of $B-V$, $R'_{HK}$, and Li EW of the stars in comparison with a subset of our benchmark clusters. We then compare the age posterior computed using {\tt BAFFLES} to ages from the $R'_{HK}$ polynomial in \citet{mamajek:2008} (“MH08 age”, though we again note that \citealt{mamajek:2008} advocate a modified $R'_{HK}$ relation incorporating additional correlations as well) and literature ages from \citet{mamajek:2012fomalhaut,konopacky:2016,milli:2017} for TW PsA, HR 2562, and HD 206893, respectively. For HR 2562 and HD 206893, age posteriors for calcium and lithium are multiplied together to find a final age.  The middle right plot demonstrates combining a posterior from {\tt BAFFLES} with the PDF from a different source, in this case the age PDF for Fomalhaut derived by \citet{nielsen:2019}. }
\label{fig:notable_stars}
\end{figure*}

TW PsA is a stellar companion to the A3V star Fomalhaut with a bright debris disk and planetary companion. The system's age has been estimated by \citet{mamajek:2012fomalhaut} to be 440 $\pm$ 40 Myr by combining independent age estimates from isochrones, rotation rate, X-ray luminosity, and lithium abundance. From lithium alone, \citet{mamajek:2012fomalhaut} estimates an age of 360 $\pm$ 140 Myr by comparing the Li EW of TW PsA, with values $B-V$=1.1 \replaced{(Simbad)}{($V$ from \citealt{keenan:1989} and $B$ from \citealt{cutri:2003})} and Li EW=$33\pm2$ \citep{barrado_y_navascu:1997}, to the Li EW in the clusters Pleiades, M34, UMa, and Hyades. Using these same values of $B-V$ and Li EW as input to {\tt BAFFLES}, we report an age of \replaced{305 Myr with a 68\% confidence interval between 257 Myr - 371 Myr}{295 Myr with a 68\% confidence interval between 213 Myr - 371 Myr} (third panel of Fig. \ref{fig:notable_stars}), consistent with the \citet{mamajek:2012fomalhaut} lithium age, but a factor of $\sim$\replaced{1.4}{1.5} too young for the final adopted age. However at $B-V$=1.1, we are limited by our cluster samples which have lithium detections up to the age of M34 (240 Myr), and non-detections at the age of Coma Ber (600 Myr), but no information in between. Thus interpolations to older ages at this $B-V$ are difficult with our current dataset.      

We also combine our age PDF with that for the A star Fomalhaut from \citet{nielsen:2019}, $750^{+170}_{-190}$ Myr, with our PDF (middle-right plot of Fig. \ref{fig:notable_stars}), to get a final age for the system, \replaced{$349^{+40}_{-68}$ Myr}{$356^{+58}_{-75}$ Myr}. Since the distribution from {\tt BAFFLES} is significantly narrower than that from \citet{nielsen:2019}, the product age changes little, yet this serves as an example of how an age posterior allows ages from {\tt BAFFLES} to be robustly combined with ages from other sources.

HR 2562 is an F5V star around which a brown dwarf companion was discovered with the Gemini Planet Imager in 2016 \citep{konopacky:2016}. \citet{Asiain:1999} estimated the age to be 300 $\pm$ 120 Myr based on space motions and evolutionary model-derived ages.  \citet{Casagrande:2011}, using Strömgren photometry and isochrones, derive a Bayesian age of 0.9-1.6 Gyr (68\% confidence interval).  From calcium alone with log($R'_{HK}$) = $-4.551$ \citep{gray:2006}, we report an age of \replaced{1200 Myr (68\%CI: 700 - 3000 Myr)}{1400 Myr (68\%CI: 690 - 3700 Myr)}.  From lithium alone, using Li EW = $21\pm5$ \citep{mesa:2018} and $B-V$=$0.45\pm.02$ \replaced{(Simbad)}{\citep{tycho_catalog:2000}}, we find an age of \replaced{0.6 Gyr (68\%CI: 0.5 - 1.3 Gyr)}{0.7 Gyr (68\%CI: 0.5 - 1.8 Gyr)}. HR 2562 is in the very center of the lithium dip, and so the depletion at this color is poorly constrained given the Hyades is the only dataset in which the dip is visible, and there are no older clusters in our sample \added{at this color}. Combining these posteriors, our final age is \replaced{680 Myr, with a 68\% confidence interval between 560 Myr - 1000 Myr}{660 Myr, with a 68\% confidence interval between 520 Myr - 1100 Myr}, consistent with the age range 300-900 Myr adopted by \citet{konopacky:2016}.  

HD 206893 is an F5V star with a brown dwarf companion inside its debris disk \citep{milli:2017}.  \citet{pace:2013}
derives its age to be 860 $\pm$ 710 Myr from chromospheric activity.  On the other hand \citet{david:2015} derive an age of 2.1 Gyr with 68\% CI between 1.2 - 4.7 Gyr using a Strömgren photometry fit to stellar atmosphere models, though given the long main-sequence lifetime of early F stars, this method is not particularly sensitive to the differences between young and intermediate ages (e.g. \citealt{nielsen:2013}).  \citet{milli:2017} adopts an age range between 200 - 2100 Myr.  Using a value of log($R'_{HK}$) = $-4.466$ \citep{gray:2006}, from calcium emission alone we derive a median age of \replaced{710 Myr (68\%CI: 400 - 2000 Myr)}{910 Myr (68\%CI: 410 - 2700 Myr)}. From lithium absorption with Li EW = $28.5 \pm 7$ m\AA\ \citep{Delorme:2017} and $B-V$=$0.44 \pm 0.02$ \citep{hog:2000tycho-catalog}, we report an age of \replaced{1.2 Gyr (68\%CI: 0.5 - 4.2 Gyr)}{1.3 Gyr (68\%CI: 0.5 - 5.5 Gyr)
}, though like HR 2562, HD 206893 is also in the center of the lithium dip. Our final age after combining these two posteriors is \replaced{550 Myr, with a 68\% confidence interval between 430 Myr - 810 Myr}{570 Myr, with a 68\% confidence interval between 380 Myr - 1000 Myr}, consistent with literature ages.

We find that {\tt BAFFLES} age posteriors for these field stars are consistent with literature ages. Both HR 2562 and HD 206893 are within the lithium dip and more data are needed to accurately map the depletion of lithium at these ages and colors. In general, however, lithium-based ages are often more constraining than calcium-based ones, given that the astrophysical scatter in $R'_{HK}$ is a more significant fraction of the total range of $R'_{HK}$. Nevertheless, the combination of these two methods tends to increase the precision on the final age posterior.

\subsection{Comparison to previous methods}

\begin{figure*}
  \centering
  \begin{tabular}{c @{\qquad} c }
    \includegraphics[page={6},width=.48\textwidth]{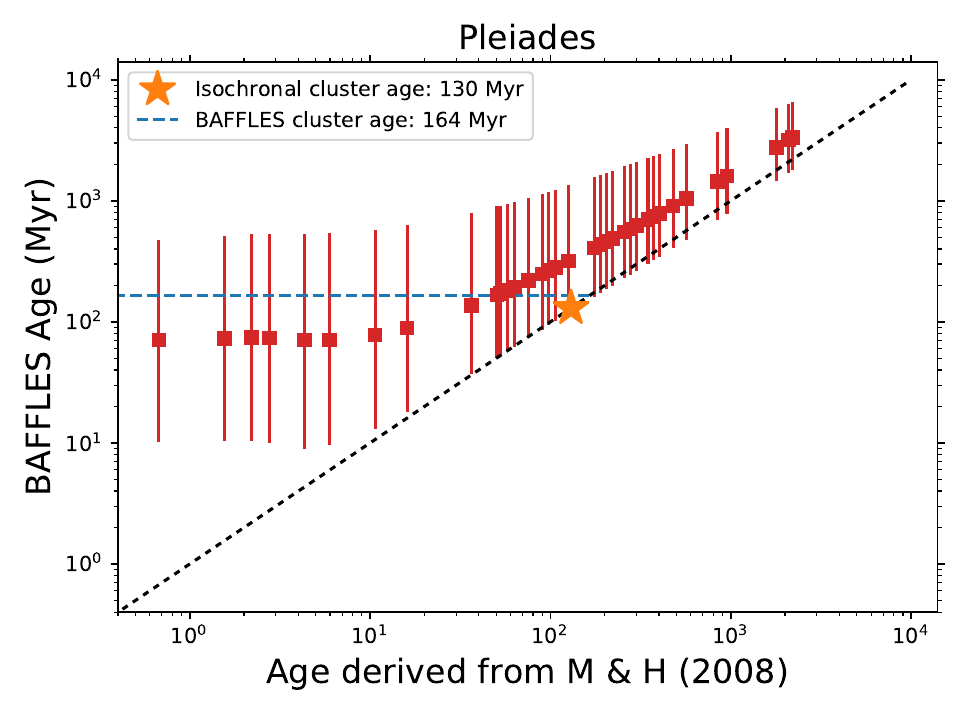} &
    \includegraphics[page={8},width=.48\textwidth]{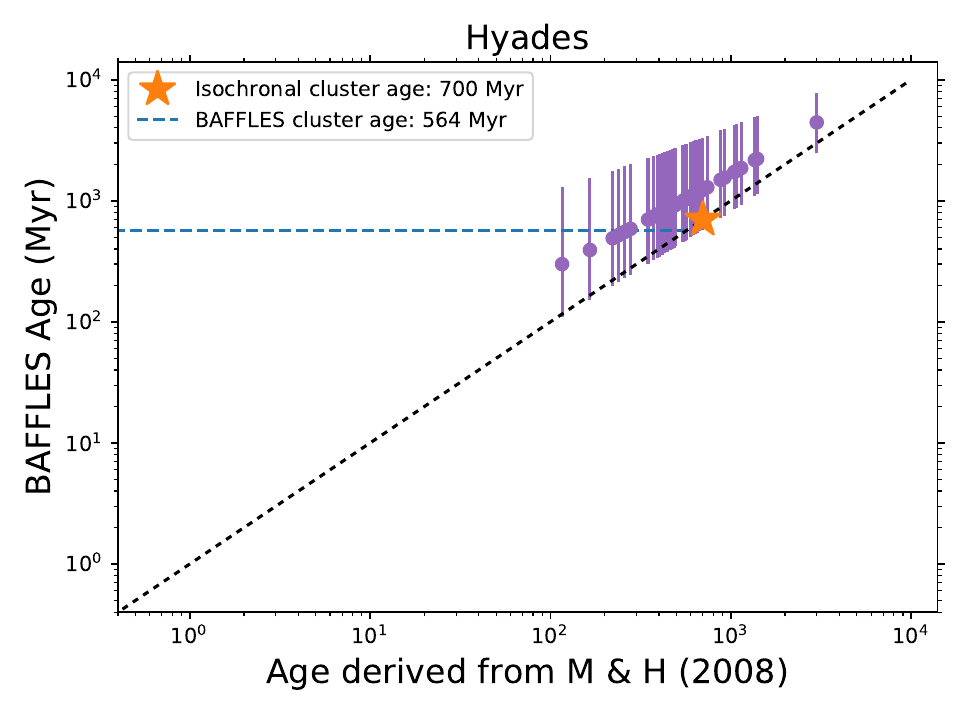} \\
  \end{tabular}
\caption{Comparison of {\tt BAFFLES} calcium age posteriors to stellar ages derived from the polynomial fit (Eq. 3) from \citet{mamajek:2008}.  Error-bars indicate the 68\% confidence interval from {\tt BAFFLES}.  As expected from the shape of the polynomial fit to $R'_{HK}$ vs time (Figure~\ref{fig:ca_vs_age}), the polynomial method tends to be biased toward younger ages. \deleted{Additionally, the larger astrophysical scatter at $\sim$100 Myr results in similar ages for the most active stars.} \replaced{Nevertheless,}{{\tt BAFFLES} produces systematically older ages for individual stars, yet} the product of these individual posteriors (Figure \ref{fig:ca_posterior_product}) shows that taken together, these posteriors are close to the correct age for the cluster as a whole. }
\label{fig:baffles_vs_mamajek}
\end{figure*}

The {\tt BAFFLES} median ages are systematically older than those derived from the \citet{mamajek:2008} $R'_{HK}$ polynomial despite relying on the same clusters and very similar fits to the clusters (Figure \ref{fig:ca_vs_age}).  In the Pleiades and Hyades, for example, the median age we derive for each star with {\tt BAFFLES} is older than the age given by the polynomial fit of \citet{mamajek:2008} (Figure \ref{fig:baffles_vs_mamajek}).  As described in Section~\ref{sec:previous_fits}, this is \replaced{a result largely}{largely a result} of the shape of the polynomial fit to mean $R'_{HK}$ as a function of time, which becomes flatter at younger ages, and so favors younger ages.  Our uniform star formation rate prior mitigates this effect, pushing each age posterior back toward older values.

\subsection{{\tt BAFFLES} ages for young, nearby stars}
We use our method, {\tt BAFFLES}, for a sample of 2630 nearby stars that appear in recent compilations of lithium measurements, $R'_{HK}$ measurements, or direct imaging surveys. In Table \ref{table:merged} we derive the ages of stars from the analysis of two direct imaging planet surveys by \citet{nielsen:2010}, from the SEEDS High-contrast Imaging Survey of Exoplanets and Disks \citep{brandt:2014_table}, from the compilation of $R'_{HK}$ values by \citet{boro_saikia:2018}, and from the lithium measurements in the spectroscopic survey of \citet{Guillout:2009li-survey}.  \citet{boro_saikia:2018} compiled $R'_{HK}$ values from a number of previous literature surveys, including \citet{Arriagada:2011,wright:2004,isaacson:2010,henry:1996,gray:2006,hall:2009,lovis:2011,bonfils:2013,duncan:1991,baliunas:1985}. \citet{Guillout:2009li-survey} acquired lithium and $H\alpha$ measurements of several hundred field stars.     

We compute age posteriors for each unique star from $R'_{HK}$ and Li EW separately, and when both are available, we multiply these posteriors to determine a final age.  For stars with multiple entries, we first compute the mean values of $B-V$, $R'_{HK}$ and Li EW  over all measurements, then use these means to find the age posteriors.

\begin{longrotatetable}
\begin{deluxetable*}{ccccccccccccccccccccccc}
\tablewidth{0pt}
\movetabledown=3cm
\tabletypesize{\tiny}
\setlength{\tabcolsep}{3pt}
\tablecaption{{\tt BAFFLES} ages for nearby stars \label{table:merged}}
\startdata
Name & RA & Dec & Sp. Type & $B-V$ & log$R'_{HK}$ & Li EW & Ref. & \multicolumn{5}{c}{$R'_{HK}$ Age} & \multicolumn{5}{c}{Li EW Age} & \multicolumn{5}{c}{Final Age} \\
& & & & & & & & \multicolumn{5}{c}{at Posterior CDF Value (Myr)} & \multicolumn{5}{c}{at Posterior CDF Value (Myr)} & \multicolumn{5}{c}{at Posterior CDF Value (Myr)} \\
& h m s & h m s & & mags & & (m\AA) & & 2.5\% & 16\% & \textbf{50\%} & 84\% & 97.5\% & 2.5\% & 16\% & \textbf{50\%} & 84\% & 97.5\% & 2.5\% & 16\% & \textbf{50\%} & 84\% & 97.5\% \\
\hline
BD+35 5152 & 00 00 35.97 & $+36$ 40 7.56 & G0 & 0.66 &  & 104 & 4 &  &  &  &  &  & 67 & 209 & \textbf{565} & 3160 & 10100 & 67 & 209 & \textbf{565} & 3160 & 10100 \\
HD 224783 & 00 00 38.12 & $-66$ 40 59.43 & G2IV/V & 0.55 & $-4.78$ &  & 3 & 1110 & 2430 & \textbf{4390} & 7730 & 11800 &  &  &  &  &  & 1110 & 2430 & \textbf{4390} & 7730 & 11800 \\
HD 224789 & 00 00 40.32 & $-69$ 40 33.48 & K1V & 0.86 & $-4.53$ &  & 3 & 249 & 630 & \textbf{1330} & 3510 & 9270 &  &  &  &  &  & 249 & 630 & \textbf{1330} & 3510 & 9270 \\
CCDM J00014+3937AB & 00 01 23.66 & $+39$ 36 38.56 & K0 & 0.82 &  & 24 & 4 &  &  &  &  &  & 349 & 637 & \textbf{1690} & 3640 & 6960 & 349 & 637 & \textbf{1690} & 3640 & 6960 \\
* 85 Peg & 00 02 10.34 & $+27$ 04 54.48 & G5VbFe-2 & 0.67 & $-4.84$ &  & 3 & 1510 & 3230 & \textbf{5620} & 8900 & 12200 &  &  &  &  &  & 1510 & 3230 & \textbf{5620} & 8900 & 12200 \\
\enddata
\tablerefs{(1) \citet{nielsen:2010}, (2) \citet{brandt:2014_table}, (3) \citet{boro_saikia:2018}, (4) \citet{Guillout:2009li-survey}}
\tablecomments{RA and Dec from Simbad online services. Table \ref{table:merged} is published in its entirety in machine-readable format.}
\end{deluxetable*}
\end{longrotatetable}

\section{Conclusion}
We have implemented a Bayesian framework, {\tt BAFFLES}, for determining the posterior probability density function on stellar age from measurements of $R'_{HK}$ calcium emission and/or $B-V$ color and Li EW lithium abundance. Importantly, {\tt BAFFLES} properly incorporates astrophysical scatter and physical priors. In developing this framework:

\begin{enumerate}
    \item We empirically determine the evolution over time of spectral indicators $R'_{HK}$ and Li EW for clusters of stars with well-characterized isochronal ages.  
    \item Using these benchmark clusters, we derive a numerical prior to derive age as a function of $R'_{HK}$ for stars with $0.45 \leq $B-V$ \leq 0.9$ and age as a function of $B-V$ and Li EW for $0.35 \leq $B-V$ \leq 1.9$. 
    \item From our tests, the method appears self-consistent and produces robust posteriors on age, though the uncertainty on ages derived from calcium may be \replaced{somewhat}{slightly} underestimated.       
\end{enumerate}

Looking ahead to future space missions, accurate ages become increasingly important.  In the next few years \textit{Gaia} is expected to discover thousands of exoplanets and brown dwarfs from measuring precise astrometry of host stars \citep{perryman:2014}.  \textit{The James Webb Space Telescope} (JWST), planned to launch in 2021, should be able to survey the nearest and youngest of these \textit{Gaia} targets to directly image the orbiting planets in the thermal infrared, where intermediate age ($\sim$100 Myr - 1 Gyr) planets have more favorable contrasts than the near infrared \citep{beichman:2019}.  Likewise the European Extremely Large Telescope (E-ELT) (e.g. \citealt{E-ELT:2016}), Thirty Meter Telescope (TMT) (e.g. \citealt{TMT:2016}), and Giant Magellan Telescope (GMT) (e.g. \citealt{GMT}) will in the near future advance our ability to directly image exoplanets.  For the next generation of telescopes, we will need stellar ages to help choose the targets for observing, since for direct imaging younger planets are more luminous and so easier to detect and characterize.  Similarly, when exoplanets are discovered, the ages of the host stars will allow mass determination for the self-luminous stellar companions. Additionally, significant evolution of planetary systems is predicted over hundreds of Myr \citep{Chiang:2002-planetevolution,ford:2008-planetevolution,Frelikh:2019-planetevolution}, and having a large number of giant planet systems with well-characterized ages will allow these predictions to be directly tested.  {\tt BAFFLES} will fill a unique role in producing robust age posteriors in a uniform way for lower-mass field stars.

\acknowledgements

We thank Eric Mamajek for helpful conversations that improved this manuscript, and for compiling ``The Lithium Plot"\footnote{\url{http://www.pas.rochester.edu/~emamajek/images/li.jpg}}, which inspired some of this work. This research has made use of the SIMBAD and VizieR databases, operated at CDS, Strasbourg, France. R.D. acknowledges support from the Fonds de Recherche du Qu\'{e}bec. Supported by NSF grants AST-1411868 (E.L.N., B.M.), and AST-1518332 (R.D.R.). Supported by NASA grants NNX14AJ80G (E.L.N., B.M.), NNX15AC89G and NNX15AD95G (B.M., R.J.D.R.).

\software{Astropy \citep{astropy}, SciPy \citep{scipy}} 

\bibliographystyle{apj}   
\bibliography{refs}

\appendix
\section{$B-V$ references}

\startlongtable
\begin{deluxetable}{ccccc}
\tablewidth{0pt}
\movetabledown=3cm
\tabletypesize{\tiny}
\setlength{\tabcolsep}{3pt}
\tablecaption{$B-V$ references for AB Dor, Tuc/Hor, and $\beta$ Pic. \label{table:B_V_ref}}
\startdata
Name & SpT & Moving Group & $B-V$ & ref \\
\hline
GSC 08894-00426 & M5Ve & AB Dor & 1.551 & 10,9 \\
HD 217343 & G5V & AB Dor & 0.64 & 2 \\
HD 218860 & G8V & AB Dor & 0.738 & 2,9 \\
HD 224228 & K2V & AB Dor & 0.985 & 7 \\
HD 35650 & K6V & AB Dor & 1.311 & 7,9 \\
HD 45270 & G1V & AB Dor & 0.602 & 7 \\
HD 65569 & F5III & AB Dor & 0.42 & 2 \\
HIP 14809 & G5 & AB Dor & 0.63 & 2 \\
HIP 17695 & M3.0V & AB Dor & 1.511 & 7,9 \\
HIP 26369 & K6Ve & AB Dor & 1.205 & 7 \\
HIP 31878 & K7V(e) & AB Dor & 1.297 & 7,9 \\
HIP 6276 & G9V & AB Dor & 0.8 & 2 \\
HR 2468 & G1/2V & AB Dor & 0.62 & 6 \\
UY Pic & K0V & AB Dor & 1.094 & 2,9 \\
V372 Pup & M1Ve & AB Dor & 1.402 & 7,9 \\
CD-53 544 & K6Ve & Tuc/Hor & 1.209 & 2,10 \\
CD-60 416 & K5Ve & Tuc/Hor & 1.0 & 2 \\
CPD-64 120 & K1Ve & Tuc/Hor & 0.807 & 2,9 \\
HD 13183 & G7V & Tuc/Hor & 0.69 & 2 \\
HD 13246 & F7V & Tuc/Hor & 0.52 & 2 \\
HD 8558 & G7V & Tuc/Hor & 0.667 & 2,9 \\
HD 9054 & K1V & Tuc/Hor & 0.91 & 4 \\
HIP 105388 & G7V & Tuc/Hor & 0.65 & 9 \\
HIP 108422 & G9IV & Tuc/Hor & 0.83 & 2 \\
HIP 1113 & G8V & Tuc/Hor & 0.756 & 2,9 \\
HIP 1481 & F8V & Tuc/Hor & 0.54 & 2 \\
HIP 16853 & G2V & Tuc/Hor & 0.6 & 2 \\
HIP 21632 & G3V & Tuc/Hor & 0.61 & 2 \\
HIP 22295 & F7V & Tuc/Hor & 0.515 & 2,9 \\
HIP 2729 & K4Ve & Tuc/Hor & 1.226 & 9 \\
HIP 30030 & G0V & Tuc/Hor & 0.57 & 2 \\
HIP 30034 & K1V(e) & Tuc/Hor & 0.805 & 2,9 \\
HIP 32235 & G6V & Tuc/Hor & 0.575 & 2,9 \\
HIP 33737 & K2V & Tuc/Hor & 1.036 & 2,9 \\
HIP 490 & G0V & Tuc/Hor & 0.6 & 1 \\
HIP 9141 & G4V & Tuc/Hor & 0.673 & 2,9 \\
TYC 7600-0516-1 & K1V(e) & Tuc/Hor & 0.898 & 2,9 \\
TYC 5882-1169-1 & K3/4 & Tuc/Hor & 1.166 & 2,9 \\
G 271-110 & M4+>L0 & $\beta$ Pic & 1.803 & 10 \\
BD+30 397B & M2 & $\beta$ Pic & 1.5 & 1 \\
BD+05 378 & K6 & $\beta$ Pic & 1.309 & 4,9 \\
PM J03325+2843 & M4+M4.5 & $\beta$ Pic & 1.542 & 10 \\
UCAC2 36944937 & M5 & $\beta$ Pic & 1.0 & 10 \\
V1005 Ori & M0 & $\beta$ Pic & 1.373 & 2,9 \\
CD-57 1054 & M0.5 & $\beta$ Pic & 1.383 & 2,9 \\
UCAC3 176-23654 & M2.9 & $\beta$ Pic & 1.49 & 10 \\
AO Men & K6.5 & $\beta$ Pic & 1.251 & 2,9 \\
HD 139084 & K0 & $\beta$ Pic & 0.803 & 4,9 \\
ASAS J164301-1754.4 & M0.6 & $\beta$ Pic & 1.36 & 3,8 \\
CD-27 11535 & K5 & $\beta$ Pic & 1.084 & 2,9 \\
HD 155555C & M4.5 & $\beta$ Pic & 1.54 & 4 \\
GSC 08350-01924 & M3 & $\beta$ Pic & 1.46 & 4 \\
CD-54 7336 & K1 & $\beta$ Pic & 0.766 & 2,9 \\
HD 161460 & K0 & $\beta$ Pic & 1.495 & 4,9 \\
HD 319139 & K6 & $\beta$ Pic & 0.79 & 2 \\
GSC 07396-00759 & M1.5 & $\beta$ Pic & 1.36 & 4 \\
PZ Tel & K0 & $\beta$ Pic & 0.878 & 2,9 \\
1SWASP J191028.18-231948.0 & M4.0 & $\beta$ Pic & 1.533 & 5 \\
CD-26 13904 & K4 & $\beta$ Pic & 1.09 & 2 \\
UCAC3 116-474938 & M4 & $\beta$ Pic & 1.56 & 10 \\
SCR J2010-2801 & M2.5+M3.5 & $\beta$ Pic & 1.5 & 10 \\
AU Mic & M1 & $\beta$ Pic & 1.423 & 9 \\
CPD-72 2713 & K7+K5 & $\beta$ Pic & 1.315 & 2,9 \\
WW PsA & M4 & $\beta$ Pic & 1.516 & 7,9 \\
TX PsA & M4.5 & $\beta$ Pic & 1.57 & 4 \\
UCAC4 494-001142 & M3.9 & $\beta$ Pic & 1.561 & 10 \\
UCAC2 16305530 & M4.5 & $\beta$ Pic & 1.58 & 10 \\
RX J0506.2+0439 & M3.8 & $\beta$ Pic & 1.52 & 10 \\
UCAC2 35242146 & M4.0 & $\beta$ Pic & 1.58 & 10 \\
UCAC3 66-407600 & M3.6 & $\beta$ Pic & 1.51 & 10 \\
HD 181327 & F6V & $\beta$ Pic & 0.46 & 2 \\
HD 35850 & F8V(n)k: & $\beta$ Pic & 0.537 & 2 \\
HIP 10679 & G2V & $\beta$ Pic & 0.59 & 1 \\
HIP 10680 & F5V & $\beta$ Pic & 0.49 & 1 \\
HIP 11437 & K7V & $\beta$ Pic & 1.18 & 1 \\
\hline
\enddata
\tablerefs{(1) \citet{1987AAS...71..413M}, (2) \citet{2000AA...355L..27H}, (3) \citet{2003AJ....125..984M}, (4) \citet{2006AA...460..695T}, (5) \citet{2006AJ....132..866R}, (6) \citet{2010AA...520A..15M}, (7) \citet{2010MNRAS.403.1949K}, (8) \citet{2011MNRAS.411..117K}, (9) \citet{2012AcA....62...67K}, (10) \citet{2012yCat.1322....0Z}}
\tablecomments{Stellar $B-V$ values and references are for stars of AB Dor, Tuc/Hor, and $\beta$ Pic moving groups, whose sources for Li EW did not include $B-V$ values. AB Dor and Tuc/Hor stars are from \citet{mentuch:2008}, and $\beta$ Pic stars are from \citet{mentuch:2008} and \citet{shkolnik:2017}. Note that a single $B-V$ reference is for both B and V magnitudes while two references are for B magnitude and V magnitude respectively. }
\end{deluxetable}

\textcolor{white}{MORE TEXT FOR FORMATTING PURPOSES. OTHERWISE TABLE DOES NOT SHOW UP.}

\newpage

\end{document}